\documentclass{aastex7}
\usepackage{placeins}
\usepackage{subfig}

\published{2026 November 19}
\submitjournal{AJ}

\begin{document}

\title{Modeling Asymmetries in Be Star Disks Using Spectro-interferometric H$\alpha$ Observations \\ from the CHARA Array’s VEGA Instrument}

\author[orcid=0000-0003-0392-1094,sname='Wysocki']{Peter Wysocki}
\affiliation{Center for High Angular Resolution Astronomy, Department of Physics and Astronomy, Georgia State University, P.O. Box 5060, Atlanta, GA 30302-5060, USA}
\email[show]{peter.wy@comcast.net}  

\author[orcid=0000-0001-8537-3583, sname='Gies']{Douglas Gies} 
\affiliation{Center for High Angular Resolution Astronomy, Department of Physics and Astronomy, Georgia State University, P.O. Box 5060, Atlanta, GA 30302-5060, USA}
\email[show]{dgies@gsu.edu}

\author[orcid=0000-0001-7425-5055,sname=Mourard]{Denis Mourard}
\affiliation{Laboratoire Lagrange, Universit\'{e} C\^{o}te d’Azur, Observatoire de la C\^{o}te d'Azur, CNRS, boulevard de l'Observatoire, CS 34229, \\ 06304 Nice Cedex 4, France}
\email{denis.mourard@oca.eu}

\author[orcid=0000-0001-9547-8290,sname=Okazaki]{Atsuo Okazaki}
\affiliation{Faculty of Engineering, Hokkai-Gakuen University, 1-1-1, Minami 26-jo nishi, Chuo-ku, Sapporo, Hokkaido 064-09236, Japan}
\email{okazaki@hgu.jp}

\author[orcid=0000-0002-8376-8941,sname=Baron]{Fabien Baron}
\affiliation{Center for High Angular Resolution Astronomy, Department of Physics and Astronomy, Georgia State University, P.O. Box 5060, Atlanta, GA 30302-5060, USA}
\email{fbaron@gsu.edu}

\author[orcid=0000-0001-5415-9189,sname=Schaefer]{Gail Schaefer}
\affiliation{The CHARA Array of Georgia State University, Mount Wilson Observatory, Mount Wilson, CA 91023, USA}
\email{schaefer@chara-array.org}

\begin{abstract}

Be stars are rapidly rotating B-type stars with outflowing decretion disks. The observational record shows that in many cases the peaks of their 
H$\alpha$ flux profile vary on long timescales. These changes are probably caused by the development of disk brightness asymmetries 
that are related to global disk oscillation modes.  Here we present spectro-interferometric observations made with the CHARA Array interferometer 
and the VEGA beam combiner that document the interferometric data across the H$\alpha$ line for the Be stars $\gamma$~Cas, $\phi$~Per, 
$\zeta$~Tau, and $\beta$~CMi. The differential interferometric visibility amplitudes and phases across the 
H$\alpha$ emission line reveal clues about the asymmetries present in their disks. 
Here we describe a model of the H$\alpha$ formation that calculates the emission intensity at spatial positions around the star and disk.  
The parameters describing the disk and the oscillation mode are optimized to match the interferometric data and contemporaneous H$\alpha$ spectroscopy. 
The analysis confirms the presence of strong disk asymmetries at the time of the observations (2020) in the cases of $\gamma$~Cas and $\phi$~Per, 
and hints that weaker oscillation modes were also present in the disks of $\zeta$~Tau and $\beta$~CMi.  
The success of this single epoch study demonstrates the promise of future interferometric programs to investigate how the 
modes precess within the disk and grow and decline over time. 

\end{abstract}



\section{Introduction}\label{sec:IntroV2}

Be stars are non-supergiant B-type stars that show emission lines in their spectra (\citealt{2013AandARv..21...69R}). 
This emission comes from an outflowing decretion disk of hot, ionized gas around the star. Be stars are rapid rotators 
with equatorial velocities exceeding approximately 75\% of the critical rotation rate. 
Many exhibit photometric and spectroscopic variations associated with nonradial pulsations \citep{2022AJ....163..226L}.
Their high rotational velocities and strong pulsations may act to eject gas into the decretion disk 
(\citealt{2001AandA...369.1058R}; \citealt{2016ASPC..506...47K}; \citealt{2025A&A...699A..82L}). 
 The disk gas nearly follows Keplerian rotation (\citealt{meilland2007first}), but viscosity helps to transfer angular momentum outwards 
 leading to a structure described by the viscous decretion disk (VDD) model of the density distribution 
 (\citealt{1991MNRAS.250..432L}; \citealt{1999AandA...348..512P}; \citealt{2002MNRAS.337..967O}).
 
The disks of Be stars are subject to instabilities that can lead to the development of asymmetric structures. 
One key contributor to gas asymmetries are global disk oscillations (GDOs). 
GDO asymmetries appear as one-armed spiral structures formed by the  gravitational quadrupole moment of the rotationally flattened Be star, 
the disk gas viscosity, and gas motions above and below the disk plane. 
These act to create high and low density regions separated by $180^{\circ}$ in azimuth 
(\citealt{1992A&A...265L..45P}; \citealt{1997A&A...318..548O}; \citealt{2000ASPC..214..409O};
\citealt{2008MNRAS.388.1372O}; \citealt{2009AandA...504..915C}; \citealt{2015A&A...576A.112E}).  

The state of the disk and its asymmetries can be observed through spectroscopic observations of the emission line profiles and long baseline interferometry. 
The H$\alpha$ profile generally shows a double-peaked emission line, characteristic of a rotating disk (\citealt{1982BASI...10..281K}). 
For an axisymmetric disk, these two peaks will be equal in height, as there is an equal amount of gas moving towards and away from the observer. 
However, observations of Be stars often show different heights for these peaks, implying that the disks are asymmetric. 
The heights of the individual peaks can change over time, creating a varying ratio of the violet ($V$) to the red ($R$) peak (\citealt{1983AandA...128..171D}). 
Long-term observations of such $V/R$ variations show that they are often cyclic in nature, with active episodes 
lasting for years and several cycles, before entering a more quiescent state (\citealt{2023A&A...678A..47B}). 
$V/R$ variations occur in approximately two thirds of Be stars (\citealt{1982BASI...10..281K}). 
These variations suggest that the asymmetries are common, transitory, and 
quasi-periodic with timescales much longer than those of Keplerian gas motions or binary orbital periods
\citep{ 1995AandA...300..163H, 2009A&A...504..929S, 2023A&A...678A..47B}.  

The GDO asymmetry is a primary source of the large $V/R$ variations observed in Be stars (\citealt{1994A&A...288..558T}; \citealt{1995AandA...300..163H}).  The one-armed spiral of the GDO progresses slowly
around the disk, moving the density enhancement from the blue-shifted side to the red-shifted side and vice-versa. 
The progression of the GDO density wave will enhance the violet or red peak depending on the location of the density enhancement relative to the observer. 

Long baseline spectro-interferometry can be used to resolve the larger features in a disk, 
and so is a powerful tool to confirm the presence and properties of a GDO asymmetry. 
For example, the differential interferometric visibility amplitude (DVA) measures the angular size at different wavelengths compared to that in the continuum, 
and the differential interferometric visibility phase (DVP) measures the photocenter displacement at different wavelengths relative to the continuum flux of the star
\citep{2012ApJ...744...19K}.  Thus, we can determine how the size and location of different regions are influenced by GDO asymmetries in the disk.

The search for disk asymmetries has focused on spectro-interferometry of the H$\alpha$ and Br$\gamma$ emission lines.  
The first detection of a disk asymmetry was made by \citet{1998AandA...335..261V} who observed a shift in the photocenter 
of the H$\alpha$ emission line of the Be star $\zeta$~Tau, and the spatial shifts corresponded 
to changes in the $V/R$ ratio of the emission line peaks between 1993 and 1994. 
Subsequently, \citet{1999AandA...345..203B} measured similar spatial shifts related to the $V/R$ variations of H$\alpha$ in $\gamma$~Cas.  
The introduction of the VLTI AMBER instrument ushered in a series of Br$\gamma$ investigations of the spatial properties 
of Be star disks \citep{meilland2007first, 2009AandA...504..915C, 2012ApJ...744...19K, 2012AandA...538A.110M, 2019AandA...621A.123C}.  
These observations were generally fit with symmetric disk models except in the cases of $\kappa$~CMa \citep{meilland2007first}, 
HD~28497 \citep{2019AandA...621A.123C}, and especially $\zeta$~Tau \citep{2009AandA...504..915C}.
\citet{2009AandA...504..915C} presented the first detailed comparison of GDO model fits with spectroscopic and interferometric observations of $\zeta$~Tau over several cycles of striking $V/R$ variability.  
\citet{2010ApJ...721..802P} obtained Br$\gamma$ and Pfund-series observations with the Keck Interferometer of the Be star 48~Lib that 
offer evidence of a disk asymmetry associated with a disk oscillation mode.

The installation of the VEGA (Visible spEctroGraph and polArimeter) instrument at the CHARA Array enabled a series of studies of the H$\alpha$ emission 
environments of Be stars \citep{2011AandA...529A..87D, 2011AandA...532A..80M, 2012AandA...545A..59S, 2015AandA...577A..51M, 2020AandA...636A.110D}.  
The VEGA measurements were fit with symmetric disk models, but two of these investigations deserve special note.  
\citet{2012AandA...545A..59S} found that symmetric disk models were adequate to fit the H$\alpha$ observations of $\gamma$~Cas, 
but they noted that the $V/R$ variations at the time of their study (2008 -- 2010) had a much smaller amplitude (and weaker GDO activity) 
compared to that at the time of the \citet{1999AandA...345..203B} observations (1988 -- 1994).  
\citet{2015AandA...577A..51M} used VEGA observations to make isovelocity image reconstructions of the H$\alpha$ emission of $\phi$~Per, 
and these showed enhanced emission from the north-west part of the disk that corresponded to a stronger red peak in the profile at that epoch. 
This provided the first interferometric evidence of GDO-related variations in the disk of $\phi$~Per. 

Here we present an analysis of long baseline interferometry of the H$\alpha$ emission in the spectra of the 
Be stars $\gamma$ Cas, $\phi$ Per, $\zeta$ Tau, and $\beta$ CMi.  
These observations were made in late 2020 with the  VEGA beam combiner at the CHARA Array 
that is readily able to resolve the circumstellar disks of these targets.  The VEGA observations provide sufficient spectral dispersion
to explore the spatial light distribution of the disks 
over a number of spectral resolution elements across the emission profile.  Each of these monochromatic 
measurements samples a specific volume of disk gas with the corresponding Doppler shift for Keplerian motion, 
so the observations are an important probe of the spatial and kinematic gas distribution in the disk.  
The observations were obtained with the three shortest baseline pairs of telescopes at the CHARA Array to explore 
the disks in two spatial dimensions in the sky.  The measurements are an important complement to studies of the 
H$\alpha$ profile shape that is inherently a one-dimensional (Doppler shift) probe of the disk structure. 
We compare the new observations with predictions from a relatively simple model of the GDO asymmetry that is incorporated into
a radiative transfer code to calculate spectral line profiles and monochromatic images of the disk in the sky.  
Section 2 describes the CHARA VEGA measurements together with contemporaneous, high spectral resolving power,
H$\alpha$ spectroscopy.  The GDO model, code, and predicted quantities are discussed in Section 3. 
Section 4 reports on the GDO model fits made of the interferometry and emission profile for each of the targets. 
Our main conclusions about the occurrence of the GDO asymmetries are summarized in Section 5.


\section{Observational Program}

We gathered both interferometric observations and H$\alpha$ profiles on four targets, $\gamma$ Cas, $\phi$ Per, $\zeta$ Tau, and $\beta$ CMi. 
These Be stars are relatively close by and have large disks, so their disks can be angularly resolved with long baseline interferometry. 
All the observations were obtained over several weeks in 2020 Nov - Dec, so they provide a snapshot of the disk state at one epoch. 
The H$\alpha$ emission was particularly strong at that time in the spectra of $\gamma$ Cas and $\phi$ Per.
Three of the targets are known spectroscopic binaries.  The companion is probably a white dwarf in the case of 
$\gamma$ Cas \citep{Naze2026} and $\zeta$ Tau \citep{Naze2024}, and
it is a faint, hot, envelope-stripped star in the case of $\phi$ Per \citep{2015AandA...577A..51M}.
These companion stars are too faint to be detected in our observations.  
Each interferometric observation gives us two observables, the differential visibility amplitude and differential visibility phase 
as a function of wavelength across the H$\alpha$ emission line. Taken together with the H$\alpha$ flux profile, we have 
three observables describing each Be star system. Global fits of all three observables are presented in subsequent sections. 

\subsection{BeSS}\label{sec:BeSS}

The H$\alpha$ profiles were obtained from the Be Star Spectra (BeSS) database\footnote{\url{http://basebe.obspm.fr/basebe/}}
\citep{2011AJ....142..149N}. The database contains a catalog of spectra of classical Be stars, Herbig Ae/Be stars, and B[e] supergiants 
that cover various wavelength regions and that were obtained by professional and amateur astronomers. 
We selected a single, high quality, H$\alpha$ spectrum of each target star that was obtained near the time of our CHARA VEGA observations. 
The details of the BeSS spectra are presented in Table \ref{tab:BeSSForModel}.  The spectral resolving power $R=\lambda / \triangle\lambda$ 
is given in column 5, and in each case, it is comparable to the GDO model value of $R=14990$ that we use below to fit the observed profiles.  
We normalized the BeSS spectra to a unit continuum by fitting a linear function to the spectral regions far from the H$\alpha$ line 
and then dividing the spectrum by this function.  Atmospheric telluric lines in the vicinity of H$\alpha$ were removed by fitting 
and dividing out the atmospheric line spectrum presented by \citet{2003AAS...203.3803H}.  Finally the spectra were transformed 
to a heliocentric wavelength reference frame.  

\begin{deluxetable*}{ccccccD}[!htb] 
\tablecaption{Selected BeSS H$\alpha$ Spectra \label{tab:BeSSForModel}}
\tablehead{
\colhead{Star} & \colhead{HD}  &  \colhead{UT}  & \colhead{BeSS}    & \colhead{$R$}& \colhead{Observer}&\twocolhead{$V_r$}          \\
\colhead{Name} & \colhead{No.} & \colhead{Date} &\colhead{File Name}& \colhead{($\lambda / \triangle\lambda $)}   & \colhead{Name}    &\twocolhead{(km~s$^{-1}$)} }
\decimals
\startdata
$\gamma$ Cas & \phn5394 & 2020 Nov 10 & \_gamcas\_20201110\_129\_Albert Stiewing.fits & 17000 & A.\ Stiewing & -13.89   \\
$\phi$ Per   &    10516 & 2020 Sep 22 & \_phiper\_20200922\_980\_34.fits              & 12000 & C.\ Buil     &   2.03   \\
$\zeta$ Tau  &    37202 & 2020 Nov 23 & \_zettau\_20201123\_933\_34.fits              & 11000 & O.\ Thizy    &  13.83   \\
$\beta$ CMi  &    58715 & 2020 Oct 31 & \_hd58715\_20201031\_132\_EBE.fits            & 14700 & E.\ Bertrand &   9.05   \\ 
\enddata

\end{deluxetable*}

\subsection{CHARA VEGA}\label{sec:VEGA}

The CHARA Array (\citealt{2005ApJ...628..453T}) is an interferometric facility operated by Georgia State University and 
is located at Mount Wilson Observatory in California. It is composed of six telescopes, 1 meter in diameter, arranged in a 
$Y$-shaped array with 15 baselines ranging from 34 to 331 meters.  
 The CHARA Array hosts several beam combiners for observations in both the near infrared and visible wavelength bands. 
The Visible spEctroGraph and polArimeter (VEGA) instrument (\citealt{2009AandA...508.1073M}) is a visible band beam combiner that was originally installed in 2007. 
It was decommissioned in 2020 December to make way for the next generation, optical-band, 6-telescope beam combiner, SPICA (\citealt{2018SPIE10701E..20M}). 
VEGA measured spectrally dispersed interference fringes, providing a limiting angular resolution of 0.3 milliarcsec (mas) for the longest baselines. 
The medium spectral resolution mode ($R = 6000$) was used in this work, and the observations were rebinned into 2 \AA\ bins spanning H$\alpha$ in order to gain in S/N ratio without much loss in spectral resolution.

\begin{figure} 
\centering
    \includegraphics[scale=1.0]{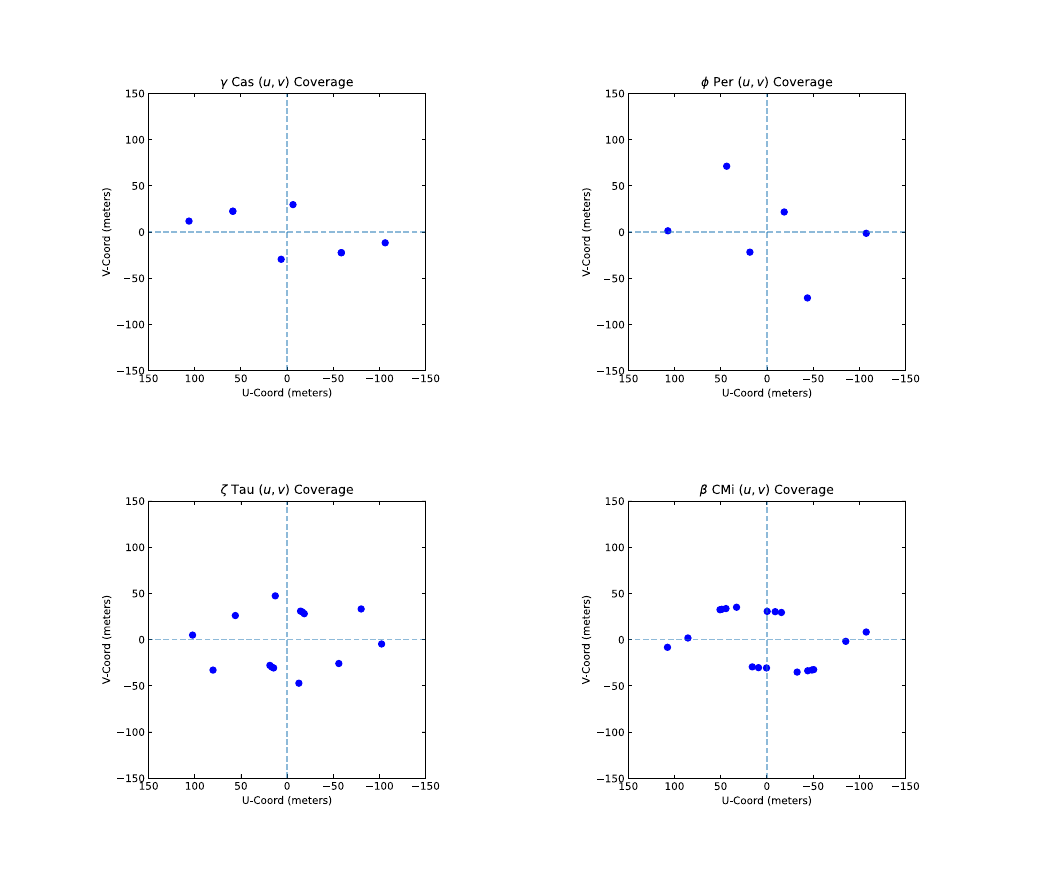}
\caption{$(u,v)$ spatial frequency coverage of the VEGA observations for $\gamma$ Cas, $\phi$ Per, $\zeta$ Tau, and $\beta$ CMi.}
\label{fig:UVPlot4}
\end{figure}

The Be star observations were obtained with VEGA in 2020 November and December in the final runs before decommissioning.  
We observed the targets using two telescopes at a time to record differential amplitudes and phases for a single baseline. 
By using different pairwise combinations of the six telescopes of the CHARA Array, we were able to obtain moderate $(u,v)$ 
spatial and orientation coverage of our target stars.  A journal of the VEGA observations is given in Table \ref{tab:TeleObs1}. 
The file name in column 1 encodes the HD catalog number of the star and the UT date and time of observation (year, month, day, hour, minute).
Column 2 gives an index number that is used in the title of the differential visibility plots below.  
Column 3 gives the CHARA telescope pair corresponding to the adjacent telescopes on the east, west, and south arms of the Array. 
Columns 4 and 5 give the projected spatial frequency coordinates as a baseline separation in meters. 
These can be converted to units of cycles per arcsec by dividing by the wavelength of each bin (in meters) and dividing by 206265 (number of arcseconds per radian).
The spatial $(u,v)$ coordinates are presented graphically for the observations in Figure \ref{fig:UVPlot4}. 
Note that two-telescope observations cover the point symmetric $(u,v)$ coordinates as well. 
For example, if one observation covered $u=+5.0$ and $v=-3.0$ m, then it also covers $u=-5.0$ and $v=+3.0$ m. 
Figure \ref{fig:UVPlot4} shows that the largest spatial frequencies appear in the east-west ($u$) direction compared to the north-south ($v$) direction.  
The largest north-south baselines were obtained with the CHARA S1-S2 telescope pair, which has the smallest baseline of the Array.  
Thus, the observations generally have better angular resolution in the east-west direction.

The VEGA data were reduced using the cross-spectrum method described by \citet{2009AandA...508.1073M}. 
Each observation was split into about 20 sequences of 25 seconds exposure each, and the best sequences were chosen for the final averages and uncertainties. 
The resulting interferometric measurements are the differential visibility amplitude and phase as a function of topocentric wavelength. 
The heliocentric velocity correction is subsequently applied to place the wavelengths in the heliocentric frame. 

\begin{deluxetable*}{cccDDD}[!htb] 
\tablecaption{CHARA VEGA Observations \label{tab:TeleObs1}}
\tablehead{
\colhead{Data File} & \colhead{Index} & \colhead{Telescope} &\twocolhead{U-COORD}& \twocolhead{V-COORD}& \twocolhead{$V_r$}\\
\colhead{Name} & \colhead{No,} & \colhead{Pair} & \twocolhead{(m)}& \twocolhead{(m)}& \twocolhead{(km~s$^{-1}$)}}
\decimals
\startdata
HD5394.2020.11.21.04.47.fits  & 1 & S1-S2 &   6.41  & -29.58 & -13.93 \\
HD5394.2020.11.22.03.38.fits  & 2 & E2-E1 & -58.72  & -22.36 & -14.00 \\
HD5394.2020.12.17.04.08.fits  & 3 & W2-W1 & 106.31  &  10.99 & -14.88 \\
HD5394.2020.12.19.01.58.fits  & 4 & E2-E1 & -58.69  & -22.53 & -14.85 \\
HD10516.2020.11.23.09.04.fits & 1 & S1-S2 &  18.56  & -21.74 &  -4.98 \\
HD10516.2020.12.16.08.46.fits & 2 & W2-W1 &  43.80  &  71.19 &   5.56 \\
HD10516.2020.12.17.04.36.fits & 3 & W2-W1 & 107.34  &   1.35 &   5.82 \\
HD37202.2020.11.21.13.08.fits & 1 & S1-S2 &  18.61  & -27.99 &  15.67 \\
HD37202.2020.11.22.12.52.fits & 2 & E2-E1 & -12.78  & -47.27 &  15.45 \\
HD37202.2020.11.23.11.15.fits & 3 & S1-S2 &  14.58  & -30.76 &  15.25 \\
HD37202.2020.11.23.12.05.fits & 4 & S1-S2 &  16.73  & -29.66 &  15.24 \\
HD37202.2020.12.16.09.28.fits & 5 & W2-W1 & 102.43  &   4.74 &  14.53 \\
HD37202.2020.12.16.12.14.fits & 6 & W2-W1 &  56.08  &  25.95 &  14.54 \\
HD37202.2020.12.17.05.26.fits & 7 & W2-W1 &  80.12  & -33.12 &  14.66 \\
HD58715.2020.11.21.10.02.fits & 1 & S1-S2 &   0.38  & -30.60 &   9.05 \\
HD58715.2020.11.21.11.56.fits & 2 & E2-E1 & -48.75  & -32.83 &   9.05 \\
HD58715.2020.11.21.13.36.fits & 3 & S1-S2 &  15.79  & -29.42 &   9.05 \\
HD58715.2020.11.22.13.18.fits & 4 & E2-E1 & -32.87  & -35.06 &   9.05 \\
HD58715.2020.11.23.11.39.fits & 5 & S1-S2 &   9.04  & -30.28 &   9.05 \\
HD58715.2020.11.24.11.29.fits & 6 & E2-E1 & -50.66  & -32.41 &   9.05 \\
HD58715.2020.11.24.12.14.fits & 7 & E2-E1 & -44.28  & -33.64 &   9.05 \\
HD58715.2020.11.24.13.00.fits & 8 & E2-E1 & -35.42  & -34.80 &   9.05 \\
HD58715.2020.12.16.09.56.fits & 9 & W2-W1 & 107.56  &  -8.26 &   9.05 \\
HD58715.2020.12.16.12.38.fits &10 & W2-W1 &  85.43  &   1.81 &   9.05 \\
\enddata
\end{deluxetable*}

In the next section, we describe model predictions for H$\alpha$ that are calculated in the rest frame, so we need 
estimates of the stellar radial velocities in order to shift the models to the observed heliocentric frame. 
$\gamma$ Cas, $\phi$ Per, and $\zeta$ Tau are known spectroscopic binaries, so their radial velocity varies with their orbital motion. 
We calculated the predicted radial velocities at the times of the observations using published orbital solutions. 
We used the radial velocity studies by \citet{2022AandA...664A.184R} for $\gamma$ Cas, 
\citet{2015AandA...577A..51M} for $\phi$ Per, and \citet{2022MNRAS.516.3366N} for $\zeta$ Tau. 
The binary status of $\beta$ CMi is controversial
\citep{2017ApJ...836..112D, 2019ApJ...875...13H, 2023Galax..11...83M}
so we simply adopted its mean radial velocity 
from the work of \citet{2019ApJ...875...13H}.  These adopted radial velocities are given in the final
column with header title $V_r$ in both Tables \ref{tab:BeSSForModel} and \ref{tab:TeleObs1}.

\FloatBarrier
\newpage


\section{Global Disk Oscillation Model}\label{sec:GDOModel}

We used an analytical radiative transfer code to model the Be star and disk system. 
This code calculates the spectral intensity produced by the central star and the circumstellar disk as functions of wavelength and spatial position.
It follows upon work by \cite{2000AandA...359.1075H} and \cite{2007ApJ...654..527G} (see especially sections 3.1, 3.2, and 4.1 in the latter). 
The main modification to this code for this work is the addition of disk density perturbations introduced by the GDO asymmetry. 
The model calculates the fractional density perturbations as functions of radial distance from the star and azimuthal position within the disk.
 
Our goal here is to model the essential GDO properties through analytical representations of the density perturbations as a function of the 
cylindrical radius from the star $x=R/R_s$ and the equatorial azimuthal coordinate $\phi$.  \citet{2008MNRAS.388.1372O} presents such 
an approximate, secular theory in his Appendix A. The disk gas is assumed to be isothermal, inviscid, and in nearly-Keplerian orbital motion.  
The deformation of the central star is ignored (setting to zero the corresponding rotational moment parameter $Q$). In this special case, the velocity perturbation 
in the radial component $u$ has an analytical solution for the $x$ dependence of the GDO eigenfunction. The solution is a modified Bessel 
function of the second kind with an imaginary order (\citealt{2008MNRAS.388.1372O}, eq.\ 22).  For a power law density distribution of 
$\rho \propto r^{-3.5}$, an integrated disk surface density $\Sigma \propto r^{-2}$, and the 3D theory of gas motion, the derived Bessel function 
imaginary order is $8i$.  This function oscillates around zero for small argument $z$, but after a final zero-crossing at $z=4.802$ the function attains 
a local maximum followed by an exponential decline (see Fig.~1 in \citealt{2008MNRAS.388.1372O}).
\citet{2008MNRAS.388.1372O} scales this last zero to the stellar surface position $x=1$, so that the mode perturbation grows from there.   
The fractional perturbation in density $s(x)$ is then derived from $u(x)$ using the solution of eq.\ A7 from \citet{2008MNRAS.388.1372O}. 
Ogilvie refers to this function as $h/c_s^2=\rho^\prime /\rho$ where $c_s$ is the isothermal sound speed and $\rho^\prime$ and $\rho$ are 
the perturbed and unperturbed gas densities, respectively.
 
\begin{figure}  
\centering
    \includegraphics[scale=0.8]{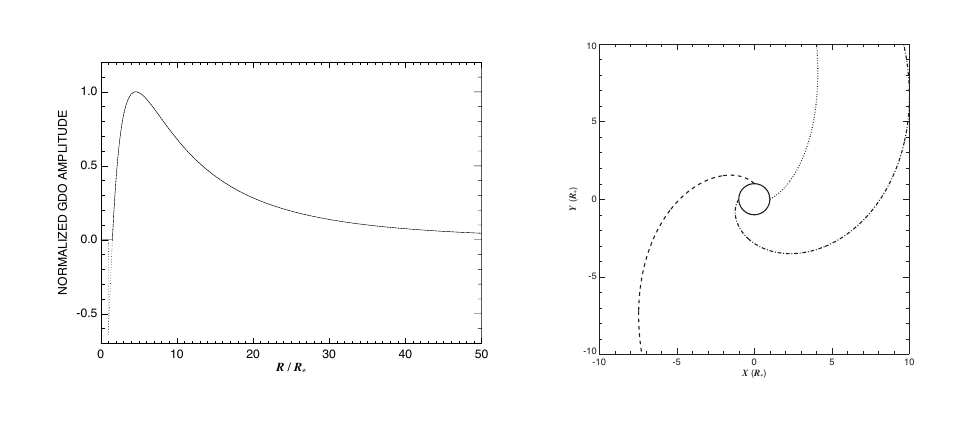}
\caption{Left: the normalized, fractional density perturbation $s(x)$ in the GDO model 
as a function of disk radial position $x=R/R_s$. The dotted line shows the functional 
reversal close to the star, and the solid line shows the adopted approximation that sets the
function to zero for $x=R/R_\star <1.53$.
Right: the geometric location of the 
density perturbation maximum in the disk plane. This map represents the spiral 
shape seen from above the disk plane.  The disk gas rotates counter-clockwise, and 
the observer is in the $X=0$ plane towards negative $Y$. The three curves show 
the spiral peak density locations for GDO parameters $\phi_0 = 0.00, 1.57, 3.14$ radians
and $a_S = 0.5, 1.0, 1.5$ as dotted, dashed, and dot-dashed lines, respectively.
\label{fig:h}}
\end{figure}

We implemented this simple approach to derive a description of the model density perturbation as a function of radial position $x$. 
We calculated a default modified Bessel function of the second kind with imaginary order $8i$ 
over a grid of $z$ values using the Fortran code DKIA developed by \citet{Gil2004}.  
Then the associated perturbation functions $u$ and $h$ were derived using the relations given by \citet{2008MNRAS.388.1372O}.  
The $u$ function is identical in appearance to that shown by the solid line in Figure~4 of \citet{2008MNRAS.388.1372O}. 
The $h$ function has a similar appearance but has a zero crossing at small $x=1.53$ (see left panel of Figure \ref{fig:h}). 
We suspect that rapid rotation, tidal forces, and mass ejection processes very close to the star probably modify or 
remove this reversal in reality, so we set the function to zero for $x<1.53$ for simplicity.  
This approximation has a negligible impact on the appearance of the model H$\alpha$ profiles.
We define the normalized perturbation function $s(x)$ by dividing by the maximum value of $h(x)$.

The maximum of the density perturbation function occurs at $x=x_{\rm max}=4.6$, but this representation is somewhat arbitrary.  
\cite{2008MNRAS.388.1372O} shows that solutions including the stellar quadrupole moment can shift the function inwards 
(more confined) while numerical experiments by \citet{2000ASPC..214..409O} and \citet{2015A&A...576A.112E} show 
that increasing viscosity acts in the opposite sense to make the function more extended (less confined). 
We explored models with rescaled eigenfunctions and found that the radial distance to the peak had only a 
minor influence on the model profiles and interferometric visibilities.  Consequently, the models that follow 
are based upon the simple radial function shown in Figure \ref{fig:h}.
 
There is no particular relation available to describe how the azimuthal position of the density perturbation maximum 
varies with radial distance $x$, because the dependence is related to assumptions about the gas viscosity
\citep{2000ASPC..214..409O, 2015A&A...576A.112E}.  However, numerical simulations indicate that the 
form is approximately a logarithmic spiral given by 
\begin{equation}
\label{eq:logspiral}
\phi_{max} (x) = a_S \ln{x}
\end{equation}
where $\phi_{max} (x)$ is the $\phi$ coordinate of the spiral perturbation maximum in cylindrical coordinates at radial position $x$
and $a_S$ is a multiplicative factor. This $\phi_{max} (x)$ position is set relative to an azimuthal offset $\phi_0$ 
of the spiral pattern at the stellar radius $x=1$ at the epoch for the model calculation.   Several examples of 
spiral patterns are shown in Figure \ref{fig:h} (right panel).  We found that $a_S \approx 1$ made a reasonable fit in most cases.  
The curvature in the spiral pattern only becomes pronounced at large $x$ where the radial perturbation function declines. 
Thus, the fits shown below were only sensitive to the value of $a_S$ in cases where the GDO amplitude is large 
($\gamma$~Cas and $\phi$~Per).  In fact, if the GDO amplitude is low ($\beta$~CMi), then the observations will only constrain  
the disk azimuth position near the peak of the radial eigenfunction at $x_{\rm max} = 4.6$ (Fig.~2). 
Then there will be a family of comparable solutions linking $a_S$ and $\phi_0$ with $a_S \ln  x_{\rm max} + \phi_0$ 
approximately equal to a constant value.

These radial and azimuthal functions for the GDO perturbation are incorporated into the main radiative transfer code 
by adjusting the local gas density at each point in the line of sight integration according to the GDO parameters and the position in the disk.  
The functional expression for the perturbed density is given by
\begin{equation}
\label{eq:Perturb}
\rho  (x,\phi,z)  = \rho_0  (x,z) ~ [1+ f ~s(x) ~\cos (\phi - \phi_{max} (x) - \phi_0) ]
\end{equation}
where $\rho_0$ is the unperturbed local disk density, $f$ is the GDO enhancement factor,  $s(x)$ is the normalized radial GDO function, 
$\phi_{max} (x)$ is the radial dependent azimuth of density maximum in the spiral pattern, and 
$\phi_0$ is the GDO azimuthal offset position at the time of observation. 
Thus, a given GDO model is dependent on only three parameters: $f$, $\phi_0$, and $a_S$.
The density enhancement factor $f$ is limited to the range from 0 to $100\%$ 
in the model application.  Model fits at the upper limit $f=100\%$ (see the cases 
of $\gamma$~Cas and $\phi$~Per below) are indicative that the density variation 
maximum may exceed the limiting value.  Thus, an estimate of a $100\%$ 
enhancement factor and its uncertainty are not physical estimates but 
signposts that the GDO amplitude exceeds the model limits.
Note that for simplicity we ignore the vertical dependence of the GDO functions, so that the fractional density perturbation 
applied above and below the disk plane is the same as in the disk plane. 
We also ignore the velocity variations induced by the GDO that are much smaller than the 
Keplerian disk velocities in the outer disk region, from which most H$\alpha$ emission arises.  
One limitation of our models is that we do not account for the continuum emission from the disk in the spectral region surrounding H$\alpha$ 
(\citealt{1988AandA...194..167D}; \citealt{2006ApJ...651L..53G}). This extra flux elevates the continuum (now star plus disk) and makes the 
H$\alpha$ line look weaker than it would be in the absence of disk continuum emission. Thus parameters related to the emission strength 
may be slightly underestimated in our models. 

The input for the simulation code includes: 
(1) a set of fixed stellar parameters (see Table \ref{tab:StarValues2} below) for an assumed spherical star;
(2) the unperturbed disk physical parameters of base density ($\rho_0$), radial density exponent ($n$), and outer radius ($R_d$);  
(3) the geometrical disk parameters of inclination ($i$) and position angle of the disk normal measured east from north in the sky ($P$);
(4) the GDO parameters of density enhancement factor ($f$), base azimuthal offset of the spiral ($\phi_0$), and the logarithmic scaling parameter for the spiral ($a_S$).  
The fitting procedure is a forward modeling scheme that optimizes the parameters from (2), (3), and (4) to create H$\alpha$ profiles and interferometric visibilities 
that match the observations. 

\begin{figure}[h] 
\centering
\includegraphics[scale=0.650]{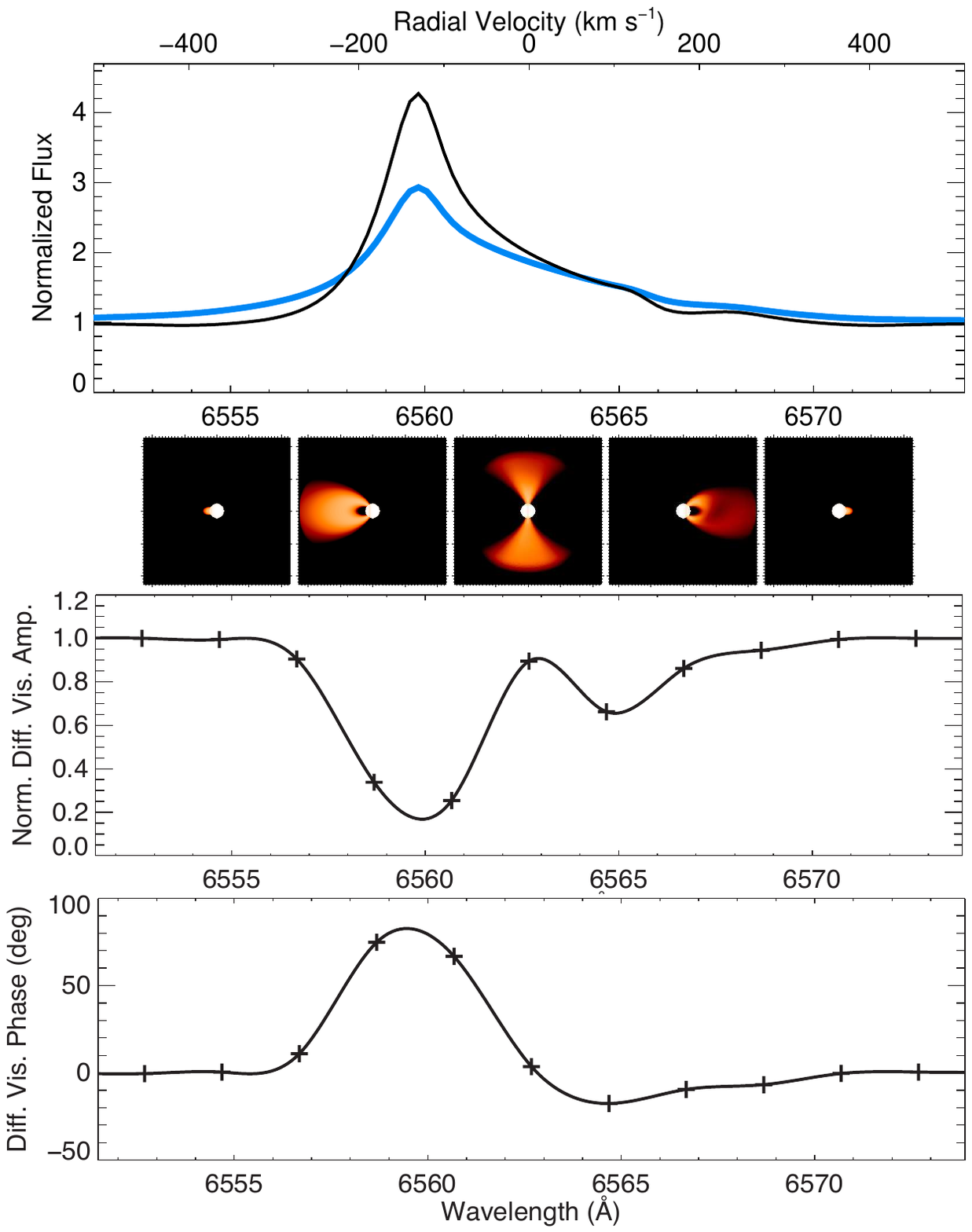}
\caption{The GDO signatures in a model simulation. 
The central five images show examples of the monochromatic intensity in the sky 
($4.5\times 4.5$ mas; logarithmic intensity scale) 
integrated over a 2 \AA\ bin in the GDO model. The disk parameters are the same as 
adopted in the model for $\gamma$~Cas (Table \ref{tab:BestFitGam}), except that the 
disk outer radius is set to $R_d =10 R_\star $ and the disk position angle is changed to 
place the projected major axis in the east-west direction, $P=0$.  Here the adopted GDO 
parameters are $f=0.9$, $\phi_0 = 1.57$ radians, and $a_S = 1.0$ (see dashed line in the right 
panel of Fig. \ref{fig:h}), which places the density perturbation maximum to the left of the star.  
The upper panel shows the resulting H$\alpha$ profile as calculated (black line) and after 
partial convolution for broadening by electron scattering (blue line). 
The panels below show the differential visibility amplitude and phase 
with the wavelength scales aligned with the central wavelengths of the images.  
These interferometric predictions were calculated for 
an observer's baseline oriented east-west with a projected separation of 50 m. 
The plus signs are the predicted values for each model wavelength bin and 
the solid lines are spline fits.  The differential visibility amplitude drops 
as the baseline samples the more spatially extended emission that is greater in
the blue-shifted part where the GDO perturbation increases the gas density. 
The differential visibility phase is related to the photocenter offset of 
the emission that again is greater on the blue-shifted side.
\label{fig:GDO_Int_Exp}}
\end{figure}

The code creates a model H$\alpha$ profile by adding the photospheric absorption component from the star with the integrated 
flux from the disk summed over the spatial pixels in the model image.  The star is assumed to be spherical and have uniform intensity. 
This simplification omits the rotational distortion and gravity darkening of the star, but the stars are essentially unresolved 
at the baselines used in this investigation (Table 3), so this omission has no impact on our analysis.  
Figure \ref{fig:GDO_Int_Exp} shows an example of several monochromatic images of the star and disk for 
a test model with a very large GDO density perturbation ($f=0.9$).  We set the 
base spiral azimuth to $\phi_0 = 1.57$ radians in this simulation in order to place the largest GDO density enhancement to the 
left of the star in the sky in the blue-shifted part of the disk (see the corresponding spiral shape in Fig. \ref{fig:h}).
The top panel shows the resulting model H$\alpha$ profile.  The profile would appear with equal height emission peaks 
near $\pm 3$ \AA\ from the rest wavelength for the unperturbed, symmetric disk case, but the introduction of the 
GDO asymmetry now skews the flux to become much stronger on the blue-shifted side and much weaker on the red-shifted side. 

Such model profiles are transformed in two ways for direct comparison with the observed profiles.  First, the models are
shifted in wavelength to the predicted stellar velocities ($V_r$ in Tables \ref{tab:BeSSForModel} and \ref{tab:TeleObs1}).
Second, we found that the wings of the model profiles were generally narrower than those of the observed profiles. 
This is a well-known problem that results from neglect of electron scattering broadening in the radiative transfer code. 
We adopted an ad-hoc scheme from \citet{2021ApJ...912...76M} to apply a broadening correction.  We suppose that 
some fraction $f_M$ of the line formation occurs behind an electron scattering screen while the remaining fraction 
$(1-f_M)$ is observed without broadening.  Then the transformed model line profile becomes 
\begin{equation}
\label{eq:Marr}
F_\lambda = (1-f_M) F_{nc} + f_M F_{c} 
\end{equation}
where $F_{nc}$ and $F_{c}$ are the non-convolved and convolved versions of the model profile, respectively. 
We used a normalized Voigt profile for the convolution function with an electron thermal velocity for the disk gas temperature 
and a fitting parameter that is the ratio of damping parameter to Doppler width, $a_V = \Gamma / (4\pi \triangle \lambda_D)$. 
Both parameters $f_M$ and $a_V$ were set by fits of the model to the observed profiles made using the non-linear, least-squares fitting code 
{\tt MPFIT} \citep{Markwardt2009}.  The blue line in the upper panel of Figure \ref{fig:GDO_Int_Exp} shows an example of the 
partial convolution for parameters $f_M = 0.56$ and $a_V = 2.1$.  

The code creates model interferometric visibility functions using the following procedure. 
The spatial intensity images described above are summed into wavelength bins of 2 \AA\ width in the same 
way as for the VEGA observations.  Then for each VEGA measurement, wavelength interpolated 
model images are created corresponding to each of the VEGA heliocentric bin wavelengths and 
the calculated stellar radial velocities.  The interferometric visibility is then calculated by making 
Fourier transforms of the images for the observed $(u,v)$ spatial frequencies using the 
method of \citet{2006ApJ...645..664A}.  
The visibility amplitude is the modulus of the resulting complex visibility. 
The visibility phase is defined by the arc-tangent of the imaginary component divided by the real component. 
The differential visibility amplitudes and phases are then set with respect to the continuum regions near H$\alpha$.
The differential amplitude is formed by dividing the visibility amplitude by the mean continuum amplitude, and
the differential phase is set by subtracting the mean continuum phase from each visibility phase.

The lower panels in Figure \ref{fig:GDO_Int_Exp} show the resulting differential visibility amplitude and phase 
for the GDO model images in the center of the figure.   These were calculated assuming the disk was observed 
with a 50~m projected baseline in the same east-west orientation as the major axis of the disk in the sky. 
The differential amplitude decreases in those wavelength bins where the disk appears larger in the sky. 
The east-west baseline sampling in this case records the largest disk size at approximately $\pm 2$ \AA\ 
from the rest wavelength in this test simulation.  The amplitude minima at these positions would be equal for 
a symmetric disk, but the GDO perturbations in this model simulation make the disk appear brighter and 
larger in the blue-shifted side compared to the red-shifted side.  Consequently, the differential amplitude 
minimum on the blue-shifted side is much deeper than that on the red-shifted side.   
The differential phase measures the photocenter displacement of the disk flux from the star in each 
wavelength bin.  In the case of a symmetric unperturbed disk, we would observe a positive shift on 
the blue-shifted side and a corresponding negative shift on the red-shifted side. However, the GDO 
perturbation now causes the shift to be much larger on the brighter, blue-shifted portion of the disk, 
creating a large phase shift at negative Doppler shift and almost no phase shift at positive Doppler shift. 
This simulation demonstrates the key diagnostic power of the differential interferometric visibility measurements. 

The value of this version of the GDO model lies in its simplicity and ease of application.  The model relies 
on just two physical parameters ($f$, $a_S$) and one geometric parameter ($\phi_0$) to represent how 
the gas density perturbations affect the line profile shape and differential visibilities.  Fits of these model 
parameters provide a useful description of the disk asymmetries in the absence of a more complete and 
detailed numerical simulation.   This apparent simplicity is the result of a number of approximations and 
rudimentary physical assumptions.  These include: (1) adopting a fixed radial functional dependence that 
ignores the stellar quadrupole moment, disk viscosity, and specific form of the disk density exponent;
(2) applying a generalized spiral shape to represent the azimuth of the density maximum as a function of 
distance from the star; (3) assuming the fractional perturbations are independent of disk normal position; 
(4) ignoring the velocity variations associated with the GDO perturbations; 
(5) making a spatially independent flux profile transformation to account for broadening by electron scattering;
and (6) omitting the continuum disk emission in the vicinity of H$\alpha$. 
Nevertheless, the GDO model presented here captures the essential properties of the disk asymmetries 
and makes model predictions that are qualitatively similar to those from numerical simulations. 
For example, the model presented by \citet{2009AandA...504..915C} uses the non-LTE radiative 
transfer code HDUST to determine how GDO perturbations develop from the gravitational potential of 
the stellar quadrupole moment.  These detailed models create a spiral density wave in the disk that is 
similar in structure to those presented here, and their model H$\alpha$ profiles and 
Br$\gamma$ differential visibilities share the characteristics of the examples we present. 
Because of the many simplifications we adopt, our model parameter fits should be regarded as 
a phenomenological representation of the GDO perturbations, and we caution that the derived 
physical properties may fail to describe the reality of the disk structure.  For example, some of the fits derived
for $\gamma$~Cas and $\phi$~Per have values of the enhancement factor that reach $100\% $. 
This implies a density of zero at one position in the disk and twice the nominal value at the point symmetric position. 
We suspect that these extremes are unrealistic and that the model results reflect some missing aspect 
of the physical state of the disk (for example, gas temperature variations caused by GDOs).  
Fully consistent physical models will be key to the analysis of future interferometric imaging of Be star disks.


\section{Modeling Results}\label{sec:ModelRes} 

Here we present fits from the GDO model of the observed H$\alpha$ flux profiles and the differential visibility 
amplitudes and phases for the particular $(u,v)$ coordinates of the VEGA measurements.   
The fits are based upon a set of adopted parameters for the Be stars given in Table \ref{tab:StarValues2}. 
These include the stellar radius, mass, effective temperature, parallax, projected equatorial rotational velocity ($V\sin i$),
and limb-darkened angular size ($\theta _{LD}$).
Table \ref{tab:StarValues2} also lists published estimates of the disk angular size 
defined by the major axis of the H$\alpha$ emission (FWHM of the wavelength integrated emission).  
These estimates demonstrate that the apparent emission disk sizes are all resolvable with CHARA VEGA.  
All four targets are the subject of prior detailed interferometric observations that were used to set the 
circumstellar disk geometry parameters of inclination ($i$)  and position angle of the disk normal ($P$).
Numerical experiments showed that these published estimates of disk orientation were fully consistent with fits 
of the VEGA interferometry observations.   The sources of the $i$ and $P$ estimates are given in the 
subsections below for each star.   The remaining parameters optimized in the fits are the disk density 
parameters ($\rho_0$, $n$, and $R_d$) and the GDO parameters ($f$, $\phi_0$, and $a_S$).

The fitting procedure was done by creating a model, deriving the predicted observables for the same wavelength bins 
and $(u,v)$ coordinates, estimating goodness-of-fit, and then changing the input parameters for a new model to improve the fit. 
The H$\alpha$ broadening parameters $a_V$ and $f_M$ are estimated by fitting each trial model to the observed flux profile 
before calculating the goodness-of-fit.   We present below fits based on three kinds of solutions for the GDO asymmetries. 

The first method zeroed-out the GDO perturbations ($f=0$) to create a symmetric disk reference solution 
to show how much the inclusion of a GDO asymmetry improves the fits.  This was done by finding the 
best fit of the H$\alpha$ profile and differential interferometric visibility amplitude (DVA) and phase (DVP) 
through optimizing the base density $\rho_0$ for a series of test values of the disk density exponent $n$. 
The $\chi^2_\nu$ statistic was calculated from residuals summed across the emission profile and over all the 
VEGA DVA and DVP observations.  This form of combined $\chi^2_\nu$ will  generally favor solutions that best fit 
the line profile because of the higher S/N and greater number of fitting points in the high resolving power spectrum.  
Then the value of $n$ yielding the lowest $\chi^2_\nu$ statistic was set for all subsequent models. 
Finally the disk outer radius $R_d$ was set to a model value that best reproduced the wavelength separation 
of the violet and red peaks of the H$\alpha$ emission line.   
The outer disk parameter $R_d$ marks the spatial limit for disk emission flux in 
the model, but the actual disk outer dimension depends upon the density power
law exponent $n$ and external factors such as a binary companion.
The disk parameters and model flux profile for this symmetric disk model are given for each star in the subsections below.

The next method relied on grid searches over the GDO parameters to find the minimum $\chi^2_\nu$ goodness-of-fit 
of  the H$\alpha$ profile and interferometric visibility.  The fit is accomplished in several grid search stages for an 
adopted value of the GDO logarithmic spiral parameter $a_S$.   The grid searches for best fit are made 
successively for models of GDO azimuthal offset $\phi_0$, base density $\rho_0$, and GDO fractional amplitude $f$.
Then the value of the spiral parameter $a_S$ that minimizes the combined $\chi^2_\nu$  is set for a final optimization of the other fitted parameters. 
Both this $\chi^2$ solution and the symmetric model fit use a wavelength range that spans the H$\alpha$ 
emission line (and is shown in the figures below).   The parameters derived by this method are used to create 
model images of the disk appearance in the sky and from above the disk plane. These parameters are also used to 
create summary plots of the VEGA DVA and DVP observations and fits. 

The final method focused on the $V$ and $R$ peaks in the emission profile rather than the entire profile. 
Historically the variations in the $V/R$ ratio provided the motivation to explore the GDO asymmetries, so 
solutions that match these peak strengths are of special interest.  Note that the peaks generally reflect the 
emission properties in the outer regions of the disk, so this criterion is dependent on the strength of the GDO 
asymmetries in the outer disk (where the GDO amplitude may be small; see Figure \ref{fig:h}). 
The $\chi^2_\nu$ statistic in this case is calculated over a restricted wavelength range that surrounds the peaks, 
and these line core ranges are indicated in the line profile figures below.  The best fits for each of the three methods 
are discussed in the next sub-sections for each Be star.. 
 
\begin{deluxetable*}{cDDcDDccc}[h] 
\tablecaption{Adopted Stellar Parameters from Literature Values \label{tab:StarValues2}}
\tablehead{
\colhead{Star}     & \twocolhead{$R_\star$}     & \twocolhead{$M_\star$}     & \colhead{$T_{\rm eff}$} & 
\twocolhead{$\pi$} & \twocolhead{$V\sin i$\phn} & \colhead{$\theta _{LD}$}  &\colhead{Disk Ang. FWHM}  & \colhead{Reference} \\
\colhead{Name}     & \twocolhead{($R_{\odot}$)} & \twocolhead{($M_{\odot}$)} & \colhead{(K)}           & 
\twocolhead{(mas)} & \twocolhead{(km~s$^{-1}$)} & \colhead{(mas)}            & \colhead{(mas)}            & \colhead{Numbers} }
\decimals
\startdata
$\gamma$ Cas &	8.12 & 14.6  & 28840 &  5.94 & \phn\phn441 & 0.45 & 3.67 & 1,2,3,4,5 \\
$\phi$ Per   &	5.7  &  9.6  & 29300 &  5.4  & 450 & 0.29 & 2.67 & 6,7,8 \\
$\zeta$ Tau  &	7.7  & 11.3  & 17950 &  7.94 & 320 & 0.57 & 3.14 & 9,10,11,12,13 \\
$\beta$ CMi  &  3.53 &  3.98 & 12871 & 20.17 & 244 & 0.66 & 2.13 & 14,15,16,17,18 \\
\enddata
\tablecomments{
(\textbf{1}) \cite{2007ASSL..350.....V},  (\textbf{2}) \cite{2001AandA...378..861C}, 
(\textbf{3}) \cite{2012AandA...545A..59S}, (\textbf{4}) J.\ Labadie-Bartz et al.\ (2026, in preparation), 
(\textbf{5}) \cite{2024MNRAS.527.2585L}, (\textbf{6}) \cite{2015AandA...577A..51M}, 
(\textbf{7}) \cite{1998ApJ...493..440G}, (\textbf{8}) This work, 
(\textbf{9}) \cite{2009AandA...504..915C}, (\textbf{10}) \cite{1997AandA...323L..49P}, 
(\textbf{11}) \cite{1990ApJS...74..595Y}, (\textbf{12}) \cite{2017IBVS.6208....1P}, 
(\textbf{13}) \cite{2024OEJV..246....1N}, (\textbf{14}) \cite{2019ApJ...875...13H}, 
(\textbf{15}) \cite{2007ApJ...654..544S}, (\textbf{16}) \cite{2012ApJ...744...19K}, 
(\textbf{17}) \cite{2015AandA...584A..85K}, (\textbf{18}) \cite{2016ASPC..506..319F}.
}
\end{deluxetable*}

\newpage
\FloatBarrier

\newpage

\subsection{$\gamma$ Cas} \label{sec:GamResults}

Our $\gamma$ Cas observations in late 2020  took place during an ``outburst'' event when the emission was very strong \citep{2023A&A...678A..47B}.
 \citet{2022AandA...664A.184R} presented a beautiful sequence of spectral profiles that document the variations seen at that time (see their Fig.~2). 
 The $V/R$ variations were especially striking in the H$\beta$ profile, and the $V/R$ ratio in this feature was at a maximum at the time of the VEGA 
 observations, indicating that a significant gas density asymmetry existed then.  Similar $V/R$ variations are present in the very strong H$\alpha$ 
 emission profile but with much lower amplitude. The H$\alpha$ line has a larger average optical depth than that for H$\beta$, so much of the 
 H$\alpha$ flux must form at larger radii (lower density) than the characteristic smaller outer radii (higher density) where H$\beta$ emission occurs.  
 The appearance of larger $V/R$ changes at smaller radii is consistent with the model prediction that the GDO amplitude is higher closer to the star 
 (see Fig.\ \ref{fig:h}). 
 
 The parameter fits for the three solution methods, $\chi^2$ minimization, $V/R$ matching, and symmetric disk (no GDO), are summarized 
 in Table \ref{tab:BestFitGam}.  
 The first three rows of Table \ref{tab:BestFitGam} give the physical parameters
 of the unperturbed disk. The adopted inclination and disk normal position angle are given in the next two rows, and these are set from 
 the recent CHARA MIRC-X results from \citet{2024ApJ...962...70K}.  These agree within uncertainties with the earlier CHARA VEGA 
 results found by  \citet{2012AandA...545A..59S}.  The next five rows give our derived estimates of the GDO parameters and the 
 H$\alpha$ broadening parameters.  
 Approximate parameter uncertainty values were estimated by calculating the goodness-of-fit over a grid of values 
 for a single parameter in the vicinity of the global best fit.   Then a conditional statistical uncertainty was set by 
 the parameter range where $\chi^2_\nu$ renormalized to a minimum of one
 was less than $1+1/{\rm dof}$ where dof is the number of degrees of freedom in the fit. 
 The final seven rows summarize the  reduced $\chi^2_\nu$ statistics and number of degrees of freedom 
 for each of the DVA, DVP, and line profile model fits.   The final $\chi^2_\nu$ statistic (marked ``All'') is the sum of the $\chi^2$ values of 
 the differential visibility amplitude and phase and the profile divided by the total number of degrees of freedom.   Columns 4, 5, and 6 list 
 the results for the fits from full line ($\chi^2$), the wavelength range close to the violet and red peaks ($V/R$), and the symmetric disk 
 model (with no GDO perturbations).  

\begin{deluxetable*}{cccccc}[!htb] 
\tablecaption{Best fit parameters for $\gamma$ Cas \label{tab:BestFitGam}}
\tablehead{
\colhead{Group} & \colhead{Parameter} & \colhead{Units} & \colhead{GDO Model ($\chi^2$)}& \colhead{GDO Model ($V/R$)}& \colhead{Symmetric Model}}
\startdata
Disk         & $R_d$ (Outer Disk Radius)            & $R_\odot$  & 675                  & 675                 & 675 \\
Disk         & $\rho_0$ (Base Density)              & g~cm$^{-3}$& $1.88\times10^{-12}$ &$1.99\times10^{-12}$ & $1.88\times10^{-12}$ \\
Disk         & $n$ (Density Exponent)               & \nodata    & 2.81                 & 2.81                & 2.81 \\
\hline
Geometry     & $i$ (Inclination)                    & deg        & 44.6                 & 44.6                & 44.6 \\
Geometry     & $P$ (Position Angle)                 & deg        & 294                  & 294                 & 294  \\
\hline
GDO          & $f$ (Enhancement Factor)             & \nodata    & $78\pm8 \% $                & $100\pm2 \% $               & 0\% \\
GDO          & $\phi_0$ (Starting GDO Angle)        & rad        & $5.36\pm0.03$                 & $3.25\pm 0.29$                & \nodata  \\
GDO          & $a_S$ (Logarithmic Spiral Factor)  & \nodata    & $1.12\pm0.16$                  & $2.71\pm0.19$                & \nodata \\
\hline
Broadening   & $a_V$ (Voigt $\Gamma / (4\pi \triangle \lambda_D)$) 
                                                    & \nodata    & 1.14                 & 1.34                & 1.05 \\
Broadening   & $f_M$ (Convolution Fraction)         & \nodata    & 0.83                & 0.85                & 0.84 \\
\hline
$\chi^2_\nu$ & Diff. Vis. Amplitude                 & \nodata    &   5.4                &   5.5               &    5.0 \\
$\chi^2_\nu$ & Diff. Vis. Phase                     & \nodata    &  27.9                &  27.4               &   33.6 \\
$\chi^2_\nu$ & Flux Profile                         & \nodata    & 797.8                & 1460.5               & 1248.4 \\
$\chi^2_\nu$ & All                                  & \nodata    & 404.2                &  265.2               &  629.0 \\
\hline
dof          & Degrees of freedom (VEGA)            & \nodata    &  63                  &  63                 &  63 \\
dof          & Degrees of freedom (Profile)         & \nodata    & 123                  &  26                 & 123 \\
dof          & Degrees of freedom (All)             & \nodata    & 249                  & 152                 & 249 \\
\hline
\enddata
\end{deluxetable*}

\FloatBarrier
\pagebreak

\begin{figure*} 
\centering
    \includegraphics[trim={0 2cm 0 2cm},clip,scale=0.85]{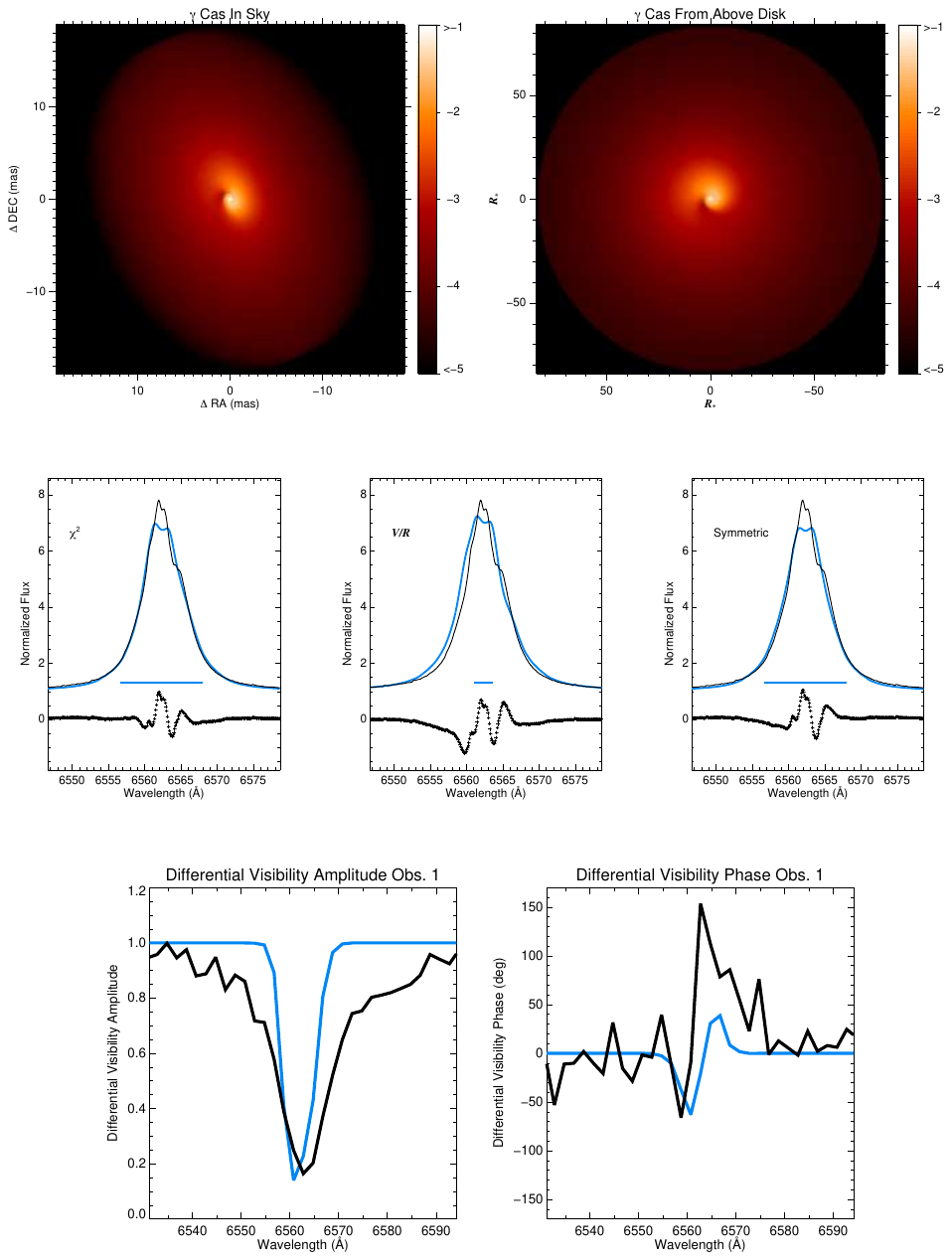}
\caption{GDO model fits for $\gamma$ Cas. 
\textbf{Top row}. Left -- Wavelength integrated (6550.68 - 6570.68 \AA ) model image of $\gamma$ Cas centered on the H$\alpha$ emission 
for  $i=44.6$ deg.  The intensity corresponds to a logarithmic scale that runs from 0.1 to 0.00001 relative to the center of the star 
(exponent marked on the side color bar).  Right -- Wavelength integrated model image for $i=0^\circ$. 
\textbf{Middle row}. H$\alpha$ profile fits for three cases.  Model results are plotted in blue and the observed profile in black. 
The horizontal blue lines below the profile indicate the wavelength range for the calculation of the $\chi^2_\nu$ goodness-of-fit statistic. 
The plus signs at bottom show the residuals (observed minus model). 
Left -- Fit obtained from the minimum of $\chi^2$. 
Center -- Fit from matching the $V$ and $R$ peak regions. 
Right -- Symmetric disk fit. 
\textbf{Bottom row}. Differential visibility amplitude and phase. Black lines mark the observations and blue lines show the minimum $\chi^2$ GDO model. 
This observation was made on 2020 Nov 21 with the S1-S2 telescope pair with spatial frequencies of $(u,v)$=$(6.41,-29.58)$ m.
\label{fig:gcfig}}
\end{figure*}

The fitting results are presented graphically in Figure \ref{fig:gcfig}.  The image in the upper left presents a wavelength integrated 
representation of the disk flux in the sky.  The intensity is normalized to the brightest pixel in the image (center of the star), 
and it is portrayed on a logarithmic scale given in the color bar.  The image on the right is the view of integrated flux from above the disk plane
(for $i=0$ and $P=0$) with a spatial scale in stellar radii.  In this format, the observer is in the $X=0$ plane in the direction of negative $Y$. 
The very strong GDO asymmetry associated with this fit ($f=78\% $) is now clearly seen in the spiral form close to the star.  The two images are 
based upon the parameters for the $\chi^2$ solution.  The middle three panels show the fits of the profile from the $\chi^2$, $V/R$, and 
symmetric disk solutions (observations in black and the model in blue).  The observed minus calculated residuals are shown as plus signs 
below the profiles.  The two plots in the lower row show the observed (black) and model (blue) differential visibility amplitude and phase. 
These differential interferometric visibility curves are associated with the first of four VEGA observations, and the full set of similar plots is given in Appendix A. 

The H$\alpha$ profile was remarkably intense in late 2020, and the $V$ and $R$ peaks were closely spaced and blended. 
The models associate these peaks with emission from the outer disk regions, and the small separation of the $V$ and $R$ peaks 
implies a very large disk outer radius where the Keplerian velocity is small.  After exploring the nominal disk parameters, 
we ended up setting the model disk outer radius to $R_d = 675 R_\odot$ that was near the maximum size to retain several pixels 
across the star in an image of $256\times 256$ pixels.  Even with this large disk outer radius, the model $V$ and $R$ peaks 
still have a larger separation than observed, indicating that the actual disk outer radius may be even larger.  
However, $\gamma$~Cas is a binary with a probable white dwarf companion \citep{2023ApJ...942L...6G, 2026A&A...707A.334N}, 
and the estimated semimajor axis is $a \approx 375 R_\odot$ \citep{2022AandA...664A.184R}.
Binary companions are expected to truncate the Be disk to a size smaller than the Be star's Roche radius
\citep{2016MNRAS.461.2616P,  2025A&A...698A.309R}, so we would expect a much smaller disk radius than indicated by the model result. 
However, disk gas may cross the gap created by the binary companion and flow into a circumbinary disk 
\citep{2016ApJ...828...47P, 2025A&A...698A.309R}.  We suspect that such a circumbinary disk existed around 
the $\gamma$~Cas system in late 2020 and that the circumbinary gas was a significant contributor to the H$\alpha$ 
emission and the closely spaced $V$ and $R$ peaks in the profile.   It is also possible that the 
small peak separation was caused by a change in the geometry of the outer disk (perhaps by a disk warping episode) 
that would reduce the inclination and the projected Keplerian velocities.  These arguments suggest that 
any structure in the outer regions of the disk resulted from the binary interaction and not the GDO perturbations 
that are mainly confined to the disk regions close to the Be star.  Thus, the GDO model predictions are probably poor
in the line core because the model ignores the binary companion and its influence on the disk structure. 

The $\chi^2$ (full profile) fit of the flux profile is significantly better than the symmetric fit, because the GDO model makes 
a better match of the line wing asymmetries (middle row of Fig.\ \ref{fig:gcfig}).  
However, the $\chi^2_\nu$ statistics for all the profile fits are large,  
as seen in the large residuals between the observed and model profiles in the line core.  The $V/R$ (profile core) model fit reaches 
closer to the observed peak height, but at the cost of a very poor match of the line wings.  We suspect that the line core mismatch 
is the result of an outer disk structure that is different from that adopted in the GDO model (see previous paragraph), 
and this discrepancy is the primary reason for the large residuals from the fit in the line core.  
The very large density enhancement factors associated with the GDO fits are probably too extreme for realistic disk oscillation modes, 
and this suggests that the simple GDO model is incomplete.  
Note that there is a third peak in the red wing of the observed profile near 6564.8 \AA\ (Doppler shift of $+112$ km~s$^{-1}$ in the stellar rest frame). 
The GDO models cannot account for narrow features like this third peak, because the radial eigenfunction and azimuthal $\cos \phi$ density 
perturbations extend over a large volume spanning a significant range in Doppler velocities.  The presence of such a narrow third peak 
suggests that additional disk structures are acting to shape the line profile.  
Indeed, this third peak morphology is seen in other Be star spectra, with explanations suggesting that they are 
caused by tidal forces from a companion \citep{2018MNRAS.473.3039P}, and/or disk warping and disk tearing caused by a 
misalignment between the plane of the disk and the companion’s orbit \citep{2024MNRAS.527.7515S, 2025MNRAS.536.2234S}.
The presence of this third peak then further supports the existence of disk structures induced by the companion that are missing from our model.

There are only four VEGA observations available of $\gamma$~Cas, and they sample only a limited range in the $(u,v)$ plane of 
spatial frequency and sky orientation (see Fig.~1).  Thus, the diagnostic power of our interferometric results is limited.  Nevertheless, the  
model fits of the differential visibility (lower panels in Figure \ref{fig:gcfig} and Appendix A) do reproduce the main observed trends but not 
the detailed curves.  The differential visibility amplitude in the VEGA observations and models show a single minimum due to the large disk size
and small peak separation of the Doppler shifts corresponding to large radii.  Note, however, that the observed differential visibility amplitude curves 
appear somewhat broader than the model curves as also found in the earlier VEGA observations by \citet{2012AandA...545A..59S}.
This may be a consequence of the neglect of electron scattering broadening in the model. 
The observed and model differential visibility phase curves show similar morphologies, but the absolute values of the phase shifts look 
larger than those in the models.  This may indicate that the GDO asymmetries have more radial extension than supposed in our simple representation. 
The large $\chi^2_\nu$ statistics of the DVA and DVP fits are due in part to these differences and to a possible underestimate 
of the observational uncertainties.  For example, the VEGA observations from 2020 Nov 22 and 2020 Dec 19 (see Fig.~8 in the Appendix) 
were made with similar $(u,v)$ coordinates, but a comparison of their DVA and DVP curves show differences that are 
larger than expected based upon the scatter in the adjoining continuum regions.  

The disk of $\gamma$~Cas is also a source of continuum flux in the vicinity of H$\alpha$.  In late 2020, the star had an apparent 
magnitude of $V=2.15$ \citep{2022AandA...664A.184R} compared to its historical faint magnitude of $V=2.6$ in 1942 when the 
disk was absent  \citep{1988Ap&SS.145..263G}.  If the difference is solely due to excess disk flux, then the ratio of disk to 
stellar monochromatic continuum flux was about 0.5 at the time of the VEGA observations, smaller than the 
ratio of 0.8 found by \citet{2012AandA...545A..59S} from VEGA observations made between 2008 and 2010. 
Any continuum disk flux will cause the H$\alpha$ emission profile to appear weaker than it would in the absence of disk flux. 
Our version of the GDO model does not include the disk continuum flux, 
so the  adopted disk base density $\rho_0$ of the model that sets the emission strength 
will be underestimated to match the weaker, diluted emission.  However, the GDO parameters for the H$\alpha$ profile set 
the fractional variations in disk density, so they are mainly unaffected by a proportional scaling of the emission strength. 
The disk continuum dilution can influence the model fits of the differential visibility.   The model DVA is normalized to a 
continuum value of one for the spatial intensity of the mainly unresolved stellar photosphere.  The observed DVA is 
normalized in the same way, but this may correspond to a partially resolved continuum disk plus star with lower 
absolute visibility.  For example, if the disk size was the same in the continuum and H$\alpha$, then the relative 
DVA would be equal to one across the profile.   We expect that the disk appears much larger in H$\alpha$ than in 
the continuum; for example, \citet{2012AandA...545A..59S} found that the continuum disk emission had 
a major axis FWHM of 0.76 mas, much smaller than the H$\alpha$ disk size.  
The disk continuum flux may cause the observed DVA to appear weaker (minima not as deep) 
compared to the model estimates calculated without the disk continuum flux.  The influence of the disk continuum 
flux on the DVP will depend upon the orientation of the GDO maximum perturbation and the $(u,v)$ coordinates of the 
observation.  A strong GDO perturbation could shift the center of light of the continuum towards the GDO density 
maximum if the disk continuum flux extends that far out from the star.  If so, then the spatial offset for the density 
perturbation maximum may be reduced in amplitude while the offset is increased for the density minimum position. 
This would probably result in a reduction in the relative asymmetry of the DVP peak and trough in cases where 
the interferometric baseline is aligned with the direction between the GDO maximum and minimum positions (see Fig.~3). 
These affects of continuum dilution may partially explain the differences between the observed and model 
DVA and DVP curves. 

\newpage

\subsection{$\phi$ Per} \label{sec:PhiResults}

The CHARA VEGA observations were obtained at a time when the H$\alpha$ emission in the spectrum of $\phi$~Per was particularly strong. 
\citet{Jonak2022} presented results of a large set of spectroscopy of $\phi$~Per that show how the emission 
strength steadily increased between 2010 and 2020.  Furthermore, there was a large $V/R$ variation then
with a cycle time of about 5 years.  The ratio of the peaks at the time of the CHARA VEGA observations 
was $V/R=0.89$ and increasing.  These H$\alpha$ variations indicate that a disk asymmetry was active at that time. 

The GDO fits for $\phi$ Per are presented in Table 5 and Figure 5 in the same format as described in the prior subsection. 
The inclination $i=77.6$ deg and disk normal position angle $P=25.7$ deg are set from the combined CHARA MIRC and 
VEGA observational study by \citet{2015AandA...577A..51M}.  
The model disk outer radius is large, $R_d = 163 R_\odot$, in order to fit the emission peak separation.  
This outer radius falls between the star's Roche radius of  $132 R_\odot$ and the binary separation of $235 R_\odot$
according to the orbital dimensions reported by \citet{2015AandA...577A..51M}.  
We normally expect that the 
disk outer radius is truncated to a size smaller than the Roche lobe \citep{2025A&A...698A.309R}, but it appears
in this case that the disk was so dense that the disk gas extended farther out towards the subdwarf companion.
Thus, we imagine that density perturbations in the outer disk were dominated by the gravity of the companion, 
and consequently the GDO model that excludes the companion will fail to represent the structure of the 
outer disk adequately.  This implies that the model fits will be poor matches of the profile in the line core that is dominated 
by flux from the outer disk. 

Both the $\chi^2$ and $V/R$ GDO model line profiles make a significantly better fit than the symmetric model 
because they better match the different peak heights (especially in the $V/R$ solution).  
The fits require a maximal GDO enhancement factor of $100\% $, implying that the disk density approaches 
zero in one part of the spiral perturbation.  This large contrast in the GDO spiral pattern is striking in  
the model image for an above-disk orientation (top right panel of Fig.\ 5).  
While we doubt that such an extreme density contrast actually existed in the disk, the fits do indicate 
that the GDO perturbation amplitude was very large at the time of the observations.
The observed H$\alpha$ profile also presents an extra peak near the $R$ peak but displaced 
towards the center of the profile, Our models cannot account for this third peak. 
We suspect that this structure in the line core is related to density changes in 
the outer disk caused by the binary companion.

\begin{figure*}[h] 
\centering
    \includegraphics[trim={0 2cm 0 2cm}, clip,scale=0.85]{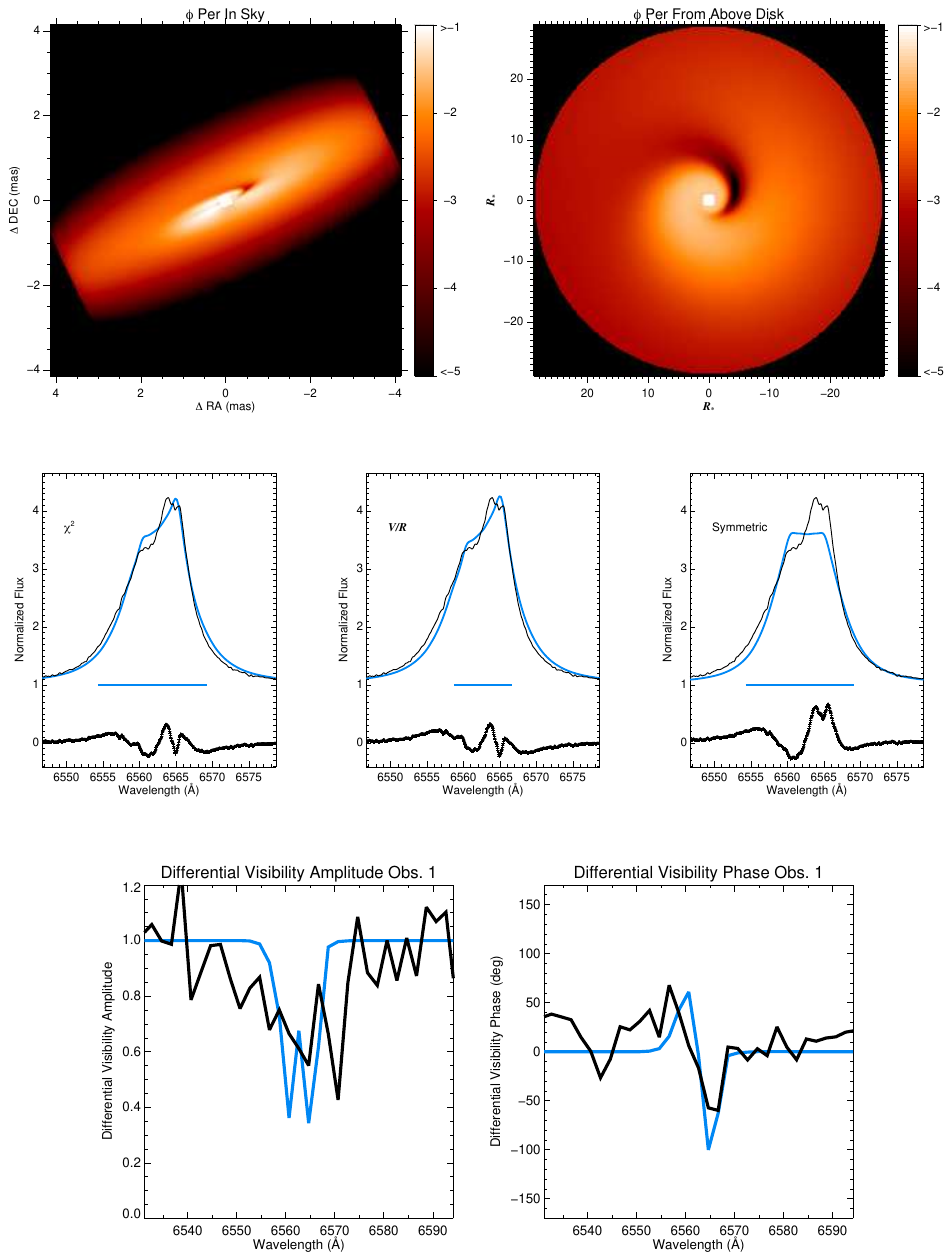}
\caption{GDO model fits for $\phi$ Per in the same format as Fig.~4. The inclination in this case is $i=77.6$ deg. 
The interferometry measurements in the bottom row are for an observation obtained on 2020 Dec 16 with the 
W1-W2 telescope pair and spatial frequencies $(u,v)$=$(43.80, 71.19)$ m. }
\label{fig:ppfig} 
\end{figure*}

The CHARA VEGA measurements for $\phi$ Per have relatively low S/N amongst our sample, and only 
three observations have high enough quality for comparison with models.  
Consequently the $(u,v)$ spatial frequency coverage is poor, and the interferometric observations
have only a limited diagnostic potential for the GDO model.
The three sets of VEGA differential visibilities (Fig.\ \ref{fig:ppfig} and Appendix B) are approximately 
matched by the model predictions.  The differential visibility amplitudes appear to have 
deeper troughs than those found for corresponding baselines in the 2011 VEGA observations 
(see Fig.\ 5 in \citealt{2015AandA...577A..51M}).  This suggests that the disk appeared larger in 2020 
compared to 2011, which is consistent with the greater emission strength in 2020 (the H$\alpha$ 
emission equivalent width changed from $-25$ \AA\ in 2011 to $-35$ \AA\ in 2020; \citealt{Jonak2022}). 
The wings of the observed DVA appear to be somewhat broader than predicted by the models, which is 
probably due to our neglect of electron scattering broadening in the model spatial Doppler maps.   

The disk is probably a significant source of continuum emission near H$\alpha$, and as discussed above (\S 4.1)
the absence of this flux component in the models can have consequences for the GDO fits. 
Analysis of earlier VEGA observations by \citet{2015AandA...577A..51M} and \citet{Jonak2022} 
suggests that the ratio of disk to star continuum flux near H$\alpha$ is in the range of 0.2 to 0.5, 
and the disk continuum flux has a projected major axis of 0.7 to 1.2 mas.   Thus, the disk 
continuum flux appears to have a much smaller spatial extent than that of the H$\alpha$ emission, 
so the omission of the disk continuum flux in the models should not affect the GDO fits significantly.   

\begin{deluxetable*}{cccccc}[h] 
\tablecaption{Best fit parameters for $\phi$ Per \label{tab:BestFitPhi}}
\tablehead{
\colhead{Group} & \colhead{Parameter} & \colhead{Units}& \colhead{GDO Model ($\chi^2$)}& \colhead{GDO Model ($V/R$)}& \colhead{Symmetric Model}}
\startdata
Disk & $R_d$ (Outer Disk Radius)                    & $R_\odot$  & 163                  & 163                 & 163 \\
Disk & $\rho_0$ (Base Density)                      & g~cm$^{-3}$& $2.64\times10^{-12}$ &$2.61\times10^{-12}$ &$2.45\times10^{-12}$ \\
Disk & $n$ (Density Exponent)                       & \nodata    & 2.68                 & 2.68                & 2.68 \\
\hline
Geometry     & $i$ (Inclination)                    & deg        & 77.6                 & 77.6                & 77.6 \\
Geometry     & $P$ (Position Angle)                 & deg        & 25.7                 & 25.7                & 25.7 \\
\hline
GDO          & $f$ (Enhancement Factor)             & \nodata    & $100\pm 2\% $                & $100\pm2 \%$               & 0\% \\
GDO          & $\phi_0$ (Starting GDO Angle)        & rad        & $0.50\pm0.03$                 & $1.13\pm 0.03$                & \nodata \\
GDO          & $a_S$ (Logarithmic Spiral Factor)  & \nodata    & $2.00\pm0.12$                 & $1.75\pm0.14$               & \nodata \\
\hline
Broadening   & $a_V$ (Voigt $\Gamma / (4\pi \triangle \lambda_D)$) 
                                                    & \nodata    & 1.73                 & 1.69                & 1.37 \\
Broadening   & $f_M$ (Convolution Fraction)         & \nodata    & 0.88                 & 0.88                & 0.92 \\
\hline
$\chi^2_\nu$ & Diff. Vis. Amplitude                 & \nodata    &   5.4                &  5.3                &   5.0 \\
$\chi^2_\nu$ & Diff. Vis. Phase                     & \nodata    &   8.8                &  9.0                &   8.9 \\
$\chi^2_\nu$ & Flux Profile                         & \nodata    & 115.0               & 82.6                & 413.2 \\
$\chi^2_\nu$ & All                                  & \nodata    &  93.4                & 58.4                & 331.5 \\
\hline
dof          & Degrees of freedom (VEGA)            & \nodata    &  38                  &  38                 &  38  \\
dof          & Degrees of freedom (Profile)         & \nodata    & 298                  & 158                 & 298  \\
dof          & Degrees of freedom (All)             & \nodata    & 374                  & 234                 & 374  \\
\enddata
\end{deluxetable*}

\FloatBarrier

\subsection{$\zeta$ Tau} \label{sec:ZetResults}

The star $\zeta$~Tau was the subject of a seminal paper by \citet{2009AandA...504..915C} who analyzed emission line profiles and VLTI AMBER 
interferometry in the context of GDO disk asymmetries.  They modeled the system using the HDUST, a 3D, non-local thermodynamic equilibrium 
code that solves the radiative transfer, radiative equilibrium, and statistical equilibrium equations. They used a vertically averaged version of 
the global disk oscillation model.  Their study focused on observations made during the period from 1997 to 2010 when the spectrum showed strong emission
and there were three cycles of marked $V/R$ variability.  Their version of the GDO model accounted for the temporal variations particularly in features
formed in the outer disk regions.  It proved more difficult to match variations of inner disk features such as the Br15 line and linear polarization. 
Their analysis was extended by \cite{2015A&A...576A.112E} who applied the 3D GDO model from \citet{2008MNRAS.388.1372O} to study the 
emission line variations.  They found that this version of the GDO model led to consistent results for the H$\alpha$ and Br15 lines of $\zeta$~Tau
by assuming a large gas viscosity to extend the GDO perturbations outwards. 

The H$\alpha$ emission strength declined significantly between 2010 and 2013 \citep{2017IBVS.6208....1P}, 
and it remained modest at least through to 2022 \citep{2022MNRAS.516.3366N}.
This decrease in strength was accompanied by the near disappearance of $V/R$ profile variability.  
\citet{2022MNRAS.516.3366N} presented a collection of representative H$\alpha$ spectra from 2017 to 2022, 
and these show that any $V/R$ variability was modest at the time of the VEGA observations in 2020 (see their Fig.\ A1).  
This record suggests that any disk asymmetries had relatively small amplitude in late 2020. 

\begin{figure*}[h] 
\centering
    \includegraphics[trim={0 2cm 0 2cm}, clip,scale=0.85]{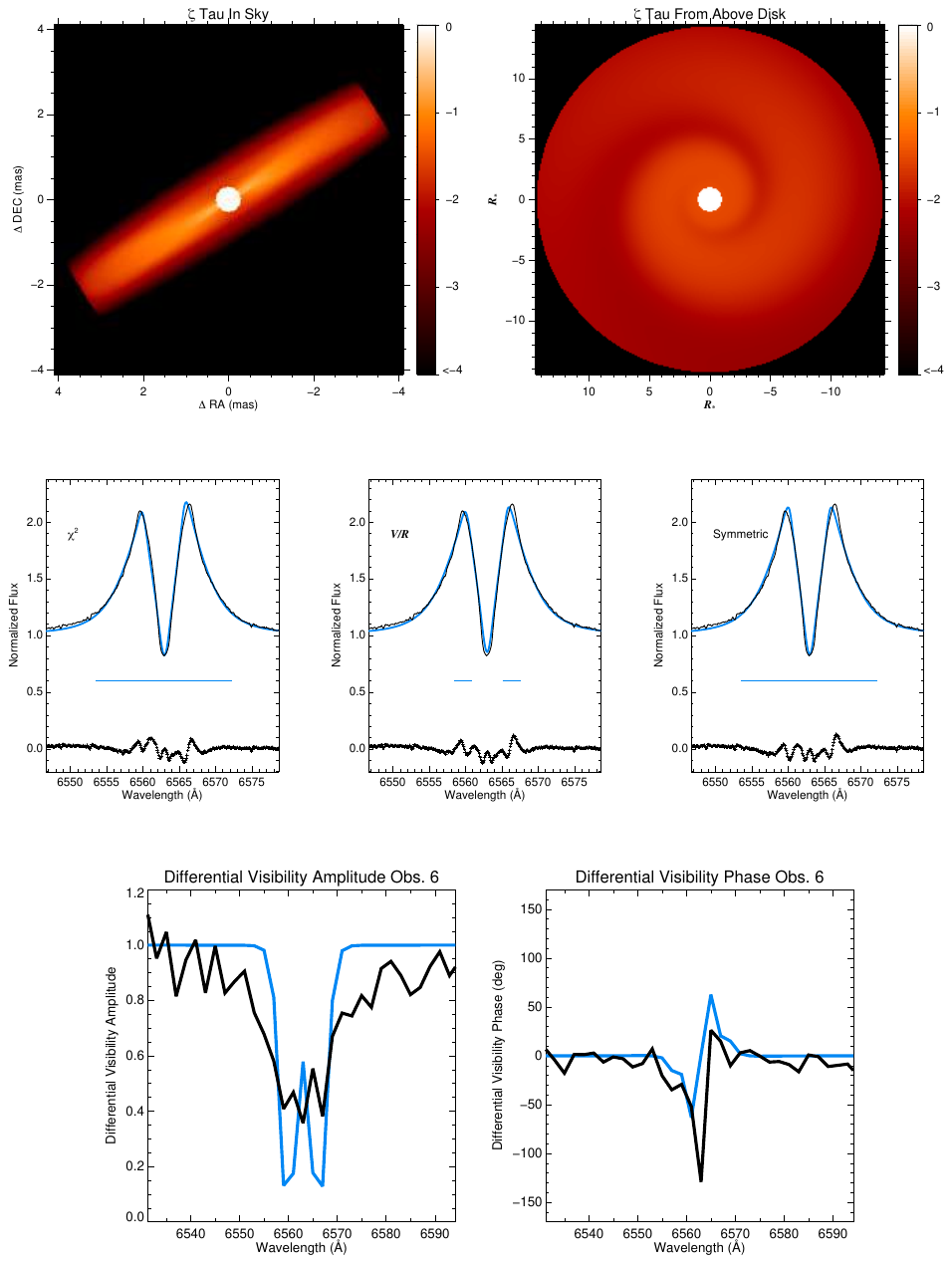}
\caption{GDO model fits for $\zeta$ Tau in the same format as Fig.~4. The inclination in this case is $i=95^\circ$. 
The interferometry measurements in the bottom row are for an observation obtained on 2020 Dec 16 
with the W2-W1 telescope pair and spatial frequencies $(u,v)$=$(56.08, 25.95)$ m. }
\label{fig:ztfig}
\end{figure*}

Our fitting results are shown in Table \ref{tab:BestFitZet} and Figure \ref{fig:ztfig}.  
The disk geometrical parameters were fixed to the estimates from \citet{2009AandA...504..915C}, 
$i = 95$ deg and $P = 212$ deg.  This places the approaching side of the disk to the west of the star 
in agreement with the original interferometric studies by \citet{1998AandA...335..261V}. 
The adopted disk outer radius $R_d = 110 R_\odot$ is smaller that the star's binary Roche radius 
of $149 R_\odot$ based upon the binary dimensions from \citet{2009A&A...506.1319R}. 
The model disk parameters are similar to those given by \citet{2009AandA...504..915C} except that 
the disk base density $\rho_0$ is about $26\times$ smaller in our fit.  This is due to the steeper drop in 
density exponent $n=3.5$ adopted by \citet{2009AandA...504..915C} compared to our result $n=2.7$
(larger $n$ requires larger $\rho_0$ for a given emission strength) and to the fact that the disk had
greater overall density in 1997--2010 compared to that in late 2020. 

Due to its high inclination, there is a 
strong shell absorption component in the center of the H$\alpha$ flux profile.
Our model can replicate this central decrease in flux very well, but like the 
other profile fits, the model line wings are not as broad as those observed. 
If we apply the default electron scattering broadening method, then we make 
a better fit of the wings but the convolution fills in the central part of 
the profile and fails to match the core absorption component. 
We dealt with this problem using a simple but ad hoc scheme. 
We extract the central absorption part of the unbroadened model 
profile and form a normalized absorption profile for the line core.  
We then multiply the broadened version from fits of the wings 
by this core absorption profile to create a modified model profile. 
The physical idea behind this transformation is that the main 
emission portion of the profile is formed in the inner part of the 
disk where the electron density is high and electron broadening 
is effective.  The absorption part, on the other hand, is mainly 
formed in the outer disk where electron scattering is minimal. 
We suppose that the outer disk acts like an absorption screen 
for disks with inclination near $i = 90^{\circ}$, so multiplying the 
broadened version (inner emission) by the absorption profile 
(opacity from outer disk) gives a first order correction of the model spectrum. 
Given the rudimentary nature of this transformation, we assigned a smaller fitting 
weight (0.1) to the pixels in the section of the profile between the peaks. 

We derive a moderate GDO density enhancement ($f=50 - 67\%$) that 
is associated with a more tightly wound spiral pattern (large $a_S$).  
This is broadly consistent with the minimal $V/R$ variability observed at this epoch \citep{2022MNRAS.516.3366N}.
The models require a non-zero GDO asymmetry in order to reproduce 
the slightly higher $R$ peak in the observed H$\alpha$ profile.  
The small GDO asymmetry results in relatively symmetrical model visibility functions,
and these approximately match the VEGA observations in some but not all cases
(Fig.\ \ref{fig:ztfig} and Appendix C).  The observed DVA  
curves are somewhat broader than those from the model that again may be due to the 
neglect of electron scattering in the calculation of the model images. 
There appears to be some inconsistency between the DVP results for the 
S1-S2 telescope pair.  The DVP plots shown in Appendix C for observation
numbers 1, 3, and 4 all were made with S1-S2 and all sample very similar spatial frequencies, 
but the sense of the DVP asymmetry is different in observation number 1. 
Thus, at least some of the blame for the relatively high $\chi^2_\nu$ for the DVP 
observations is the result of this internal inconsistency.  
Note that the large value of 
the logarithmic spiral factor $a_S \approx 4$ corresponds to a a tightly wound spiral pattern. 
This may represent a case where the GDO model parameters are being applied in a 
situation where a two-armed tidal perturbation is created by the binary companion 
\citep{2025A&A...698A.309R}.

\begin{deluxetable*}{cccccc}[h] 
\tablecaption{Best fit parameters for $\zeta$ Tau \label{tab:BestFitZet}}
\tablehead{
\colhead{Group} & \colhead{Parameter} & \colhead{Units}& \colhead{GDO Model ($\chi^2$)}& \colhead{GDO Model ($V/R$)}& \colhead{Symmetric Model}}
\startdata
Disk         & $R_d$ (Outer Disk Radius)            & $R_\odot$  & 110                  & 110                 & 110 \\
Disk         & $\rho_0$ (Base Density)              & g~cm$^{-3}$&$2.26\times10^{-12}$  &$2.30\times10^{-12}$ & $2.20\times10^{-12}$ \\
Disk         & $n$ (Density Exponent)               & \nodata    & 2.72                 & 2.72                & 2.72 \\ 
\hline
Geometry     & $i$ (Inclination)                    & deg        &  95                  &  95                 &  95 \\
Geometry     & $P$ (Position Angle)                 & deg        & 212                  & 212                 & 212 \\
\hline
GDO          & $f$ (Enhancement Factor)             & \nodata    & $50\pm4 \% $                & $67\pm3\% $               & 0\% \\
GDO          & $\phi_0$ (Starting GDO Angle)        & rad        & $3.63\pm0.14$                 & $2.08\pm0.06$                & \nodata \\
GDO          & $a_S$ (Logarithmic Spiral Factor)  & \nodata    & $4.21\pm0.14$                 & $4.91\pm0.43$               & \nodata \\
\hline
Broadening & $a_V$ (Voigt $\Gamma / (4\pi \triangle \lambda_D)$) 
                                                    & \nodata    & 2.26                & 2.26               & 2.26 \\
Broadening & $f_M$ (Convolution Fraction)           & \nodata    & 0.95                 & 0.96                & 0.95 \\
\hline
$\chi^2_\nu$ & Diff. Vis. Amplitude                            & \nodata    & 11.9                 & 12.2               & 11.7 \\
$\chi^2_\nu$ & Diff. Vis. Phase                          & \nodata    & 56.9                 & 54.6               & 63.1 \\
$\chi^2_\nu$ & Flux Profile                         & \nodata    &  8.0                 & 24.4                & 11.4 \\
$\chi^2_\nu$ & All                                  & \nodata    & 23.0                 & 31.4                & 26.1 \\
\hline
dof          & Degrees of freedom (VEGA)            & \nodata    &  86                  &  86                 &  86 \\
dof          & Degrees of freedom (Profile)         & \nodata    & 133                  &  54                 & 133 \\
dof          & Degrees of freedom (All)             & \nodata    & 305                  & 226                & 305 \\
\enddata
\end{deluxetable*}

\FloatBarrier

\subsection{$\beta$ CMi} \label{sec:BetResults}

The H$\alpha$ emission strength in the spectrum of $\beta$~CMi has shown little variability over the past three decades
\citep{1996A&AS..116..309H, 2015AandA...584A..85K, 2016ASPC..506..319F, 2019ApJ...875...13H}.  
The $V/R$ emission peak ratio shows only small ($\pm 3\% $) variations that may vary with the orbital period \citep{2023Galax..11...83M}. 
If so, such variations may be related to twin tidal arms in the disk rather than a single GDO asymmetry \citep{2025A&A...698A.309R}.
The relative quiescence of the H$\alpha$ emitting disk suggests that any GDO amplitude will be small. 

The star was the focus of a study of VLTI AMBER observations by \citet{2012ApJ...744...19K} who determined an 
inclination of $i= 38.5 \pm 1$ deg  and a position angle of the projected major axis of $137 \pm 4$ deg. 
They determined that the south-east region of the disk is approaching Earth (their Fig.\ 10). 
This orientation corresponds to a disk normal position angle of approximately $P= 45$ deg. 
These results were confirmed in a re-analysis of the VLTI AMBER data by \citet{2015AandA...584A..85K} 
and in more recent CHARA MIRC-X data obtained by \citet{2024ApJ...962...70K}. 
We adopted the latter results here  ($i= 41$ deg  and $P = 44.5$ deg). 

\begin{figure*}[h] 
\centering
    \includegraphics[trim={0 2cm 0 2cm}, clip,scale=0.85]{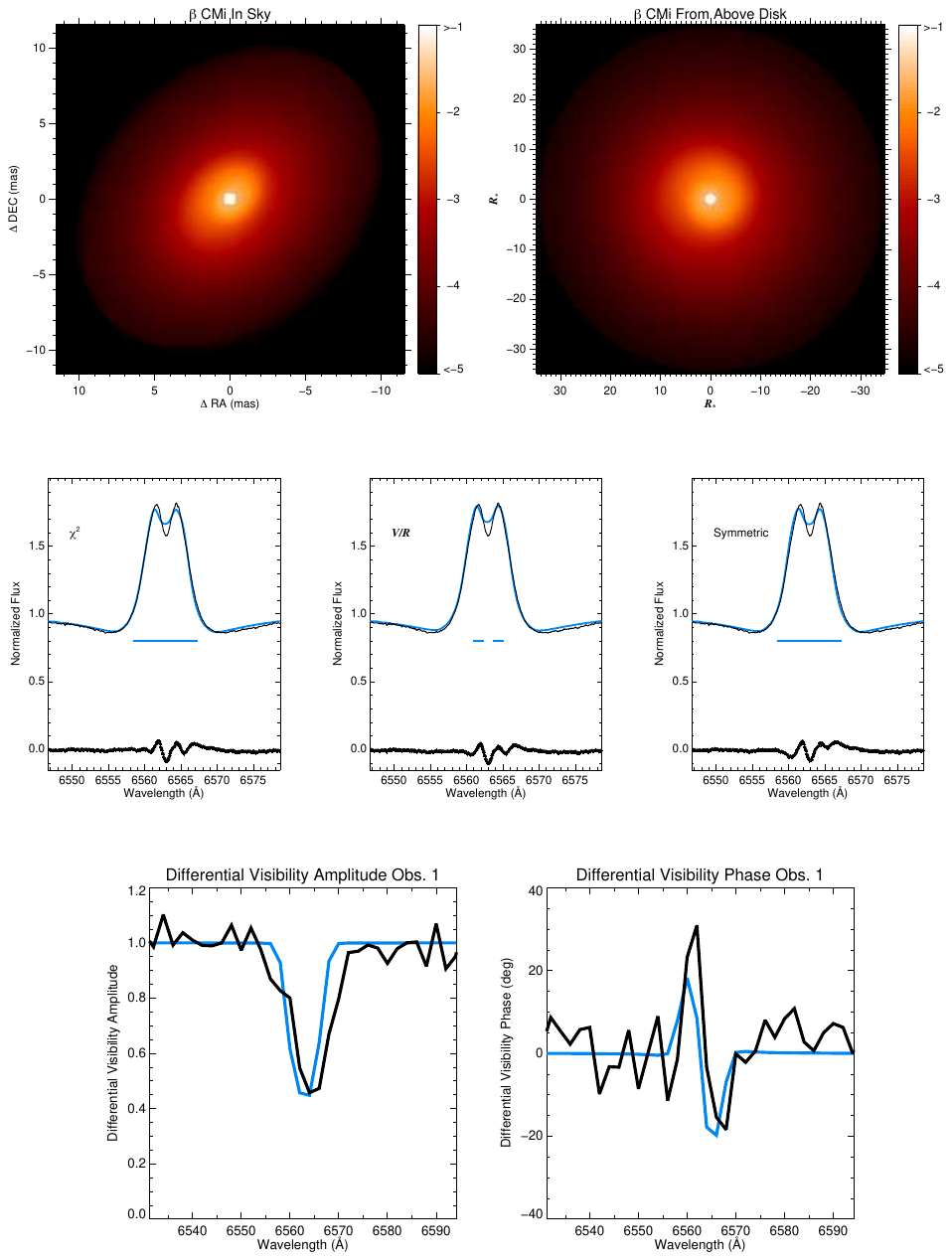}
\caption{GDO model fits for $\beta$ CMi in the same format as Fig.~4. The inclination in this case is $i=41$ deg. 
The interferometry measurements in the bottom row are for 
an observation obtained on 2020 Nov 21 with the S1-S2 telescope pair and spatial frequencies $(u,v)$=$(0.38,-30.60)$ m. }
\label{fig:bcfig}
\end{figure*}

Our fitting results are presented in Table \ref{tab:BestFitBet} and Figure \ref{fig:bcfig}. 
Our model fits for $\beta$ CMi are the best amongst all of our sample. 
Both the differential visibility amplitudes and phases (given in full in Appendix D) are fit well, 
and the fit of the H$\alpha$ flux profile is also satisfactory. 
Our derived disk base density of $\rho_0 = 2.0\times 10^{-12}$ gm~cm$^{-3}$ is in exact agreement 
with the value from \citet{2015AandA...584A..85K} 
who considered a range in the disk density exponent of $n=3.0 - 3.5$. 

The GDO amplitude is small in our fits ($f=12\% - 18\% $) so that the GDO asymmetry is hard 
to discern in the model images in Figure \ref{fig:bcfig}.
Note that in cases of small GDO amplitude, the other GDO parameters $\phi_0$ and $a_S$ 
are poorly constrained.  As noted above (\S3), in such cases there there will be a family of 
comparable solutions with $a_S \ln  x_{\rm max} + \phi_0$ equal to a constant.  
The $\chi^2$ and $V/R$ fits given in Table \ref{tab:BestFitBet} are part of such a family.
Unlike the other target cases described in the previous sections, the observed and model DVA plots 
show similar widths, suggesting that the effects of electron scattering broadening are smaller. 
This is consistent with the fits of the H$\alpha$ profile broadening for $\beta$ CMi that have the 
smallest values of $a_V$ and $f_M$ among this sample of four Be stars.  

\begin{deluxetable*}{cccccc}[h] 
\tablecaption{Best fit parameters for $\beta$ CMi \label{tab:BestFitBet}}
\tablehead{
\colhead{Group} & \colhead{Parameter} & \colhead{Units}& \colhead{GDO Model ($\chi^2$)}& \colhead{GDO Model ($V/R$)}& \colhead{Symmetric Model}}
\startdata
Disk         & $R_d$ (Outer Disk Radius)            & $R_\odot$  & 122                  & 122                 & 122 \\
Disk         & $\rho_0$ (Base Density)              & g~cm$^{-3}$&$2.01\times10^{-12}$  &$2.07\times10^{-12}$ &$2.02\times10^{-12}$ \\
Disk         & $n$ (Density Exponent)               & \nodata    & 3.38                 & 3.38                & 3.38 \\
\hline
Geometry     & $i$ (Inclination)                    & deg        &  41                  &  41                 &  41 \\
Geometry     & $P$ (Position Angle)                 & deg        &  44.5                &  44.5               &  44.5 \\
\hline
GDO          & $f$ (Enhancement Factor)             & \nodata    & $18\pm2 \% $                & $12\pm3 \% $               & 0\% \\
GDO          & $\phi_0$ (Starting GDO Angle)        & rad        & $0.93\pm0.09$                 & $4.77\pm0.06$                & \nodata \\
GDO          & $a_S$ (Logarithmic Spiral Factor)       & \nodata    & $3.37\pm0.06$                 & $1.42\pm0.44$              & \nodata \\
\hline
Broadening   & $a_V$ (Voigt $\Gamma / (4\pi \triangle \lambda_D)$) 
                                                    & \nodata    & 0.62                 & 0.92                & 0.71 \\
Broadening   & $f_M$ (Convolution Fraction)         & \nodata    & 0.64                 & 0.58                & 0.60 \\
\hline
$\chi^2_\nu$ & Diff. Vis. Amplitude                           & \nodata    &  6.4                 &  6.8                &  6.9 \\
$\chi^2_\nu$ & Diff. Vis. Phase                          & \nodata    &  8.9                 &  8.9               &  8.4 \\
$\chi^2_\nu$ & Flux Profile                         & \nodata    & 25.6                 &  8.6                & 37.6 \\
$\chi^2_\nu$ & All                                  & \nodata    & 18.7                 &  8.1                & 26.0 \\
\hline
dof          & Degrees of freedom (VEGA)            & \nodata    &  69                  &  69                 &  69 \\
dof          & Degrees of freedom (Profile)         & \nodata    & 217                  &  75                 & 217 \\
dof          & Degrees of freedom (All)             & \nodata    & 355                  & 213                 & 355 \\
\enddata
\end{deluxetable*}

\FloatBarrier


\section{Conclusions}\label{sec:Conc}

This study of high quality spectroscopy and spectro-interferometry of the H$\alpha$ emission line formed in the circumstellar disks of Be stars 
has confirmed the presence of asymmetric flux distributions of the disk gas. Such asymmetries have long been suspected to exist from observations 
of the differences in the violet ($V$) and red ($R$) flux peaks formed in the rotating disks. These spatial asymmetries as a function of Doppler shift were 
interpreted through a model of emission line formation that introduces disk gas density variations in the context of the Global Disk Oscillation model 
and that calculates the predicted line profile shape and differential interferometric visibility amplitude and phase functions.  

A comparison of the model predictions and CHARA VEGA observations from 2020 Nov and Dec show that large amplitude density asymmetries 
were present in the circumstellar disks of the Be stars $\gamma$~Cas and $\phi$~Per.  The disk H$\alpha$ emission was remarkably strong at 
this time in both stars implying that their disks were particularly dense and radially extended.  In fact, the outer disk radii adopted in our models 
of these two stars are similar to or larger than the binary separation suggesting that the disks had a circumbinary component at the time of observations. 
Contemporaneous spectroscopy documented variations in the $V$ and $R$ flux peaks that were changing on a timescale of a year or so.  
These cyclic variations probably result from the progression of the GDO mode through the disk.  
The large asymmetry found in the disk of $\gamma$~Cas in 2020 during a time of maximal emission 
stands in contrast to its absence in VEGA observations from 2008 -- 2010 when the emission was much weaker \citep{2012AandA...545A..59S}. 
The disk of $\phi$~Per was also particularly active in 2020, and the VEGA observations were made as the $V/R$ ratio was increasing from 
$V/R < 1$ to $V/R > 1$.   The GDO model fit required a very large mode amplitude to match the overall H$\alpha$ profile morphology. 

The fits of the profiles and differential visibilities of $\zeta$~Tau and $\beta$~CMi indicated marginal GDO asymmetries.  
Both targets had lower emission strength and displayed only small $V/R$ variations at that time.  This represents a very different state 
for $\zeta$~Tau that displayed stronger emission and large amplitude $V/R$ changes in the period 1997 -- 2010, the epoch 
when the emission profiles and interferometric visibilities were successfully modeled as GDO asymmetries by \citet{2009AandA...504..915C}
and \citet{2015A&A...576A.112E}.  Based on this very limited sample of Be stars, we suggest that the occurrence of GDO asymmetries 
is favored when the disk density is relatively high. 

Our model fits of the profile shapes and differential visibility functions are only partially successful in matching the observations. 
We suspect that the discrepancies are due to the simple idealized model of the GDO density perturbation adopted in the code
and the neglect of other potentially important physical processes. 
For example, the code neglects both mode dependence on position outside of the disk plane and the departures from Keplerian velocity
that are introduced by the GDO mode.  Furthermore, the model is missing the tidal influence of a binary companion that must 
play a role in the formation of a two-armed spiral perturbation as the disk outer radius becomes comparable to the binary separation
\citep{2016MNRAS.461.2616P,2025A&A...698A.309R}.
Both the H$\alpha$ profiles of the binary systems $\gamma$~Cas and $\phi$~Per display a third peak that 
suggests the existence of gas structures more complex than given by the simple GDO model.  Future interferometric studies 
with higher spectral dispersion of H$\alpha$ (CHARA SPICA) and Br$\gamma$ (VLTI GRAVITY) hold the promise of detecting 
the location and origin of such complex structures in the disk.   More extended time series of interferometric observations are required 
to confirm the expected slow
prograde rotation of the modes and to explore the timescales and geometry of their growth and dissipation. 
Multiple baseline observations that fully sample the spatial frequency $(u,v)$ plane will eventually yield spectrally resolved, monochromatic images of 
the disks that will be the ultimate test of the GDO density perturbations \citep{2015AandA...577A..51M}. 

The importance of this work is not limited to Be star disks.  We encounter gas disks in many areas of astrophysics involving accretion processes 
including proto-planetary disks, accretion disks in interacting binaries, and vast disks surrounding super-massive black holes in Active Galactic Nuclei.  
Global disk oscillation modes probably develop in many of these disk environments.  The Be star disks provide examples of such oscillation modes in 
objects that can be angularly resolved through long baseline interferometry and that vary on human timescales.  Analyses like those presented here 
of their global disk oscillation asymmetries will provide the means to understand the physical processes that drive disk instabilities throughout astrophysics.

\FloatBarrier


\newpage

\appendix

\section{$\gamma$ Cas Interferometric Fits}

\begin{figure}[h] 
\centering
\subfloat[2020 Nov 21, S1-S2, $(6.41,-29.58)$]{
  \includegraphics[scale=0.3]{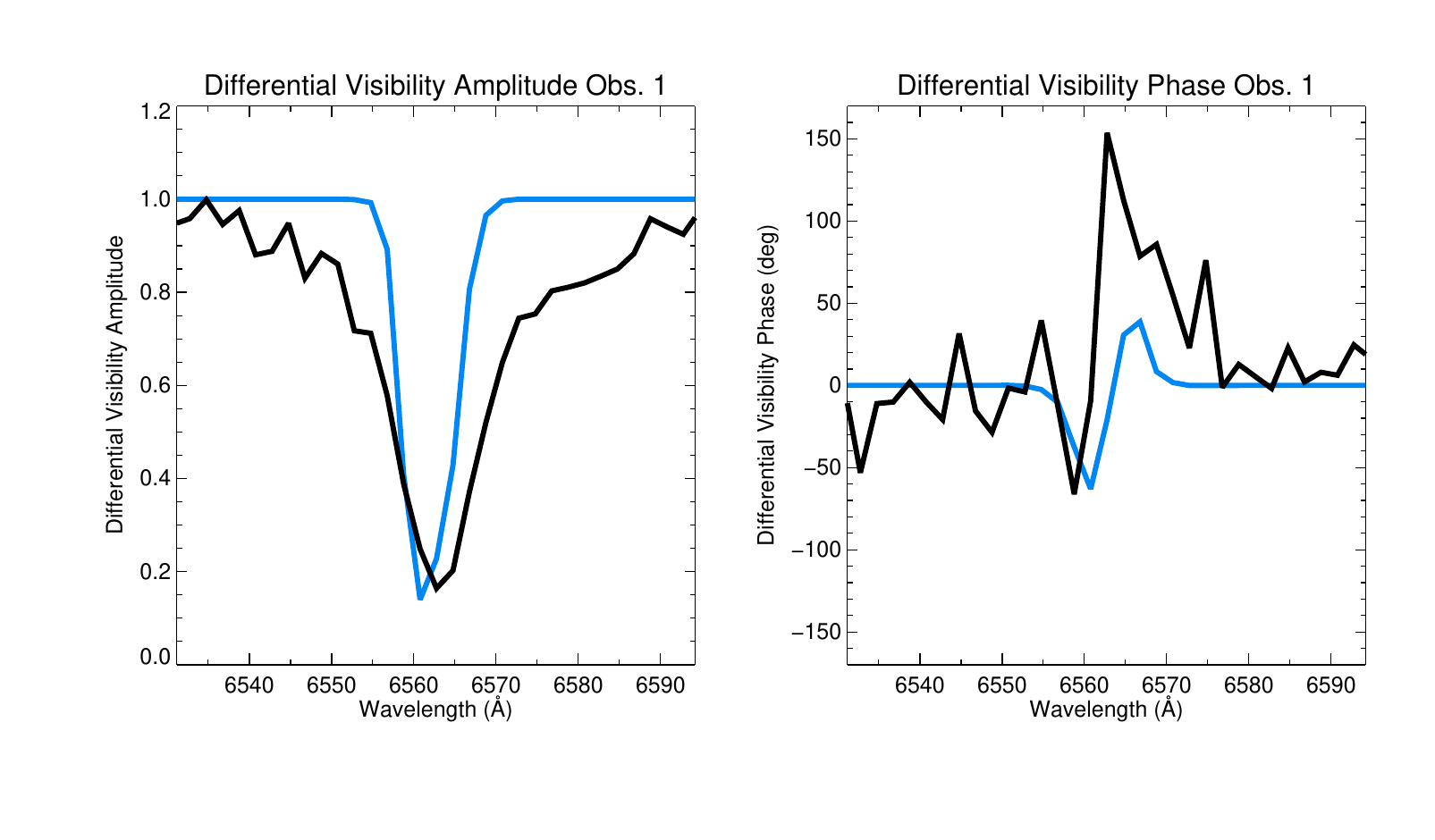}
}
\subfloat[2020 Nov 22, E2-E1, $(-58.72,-22.36)$]{
  \includegraphics[scale=0.3]{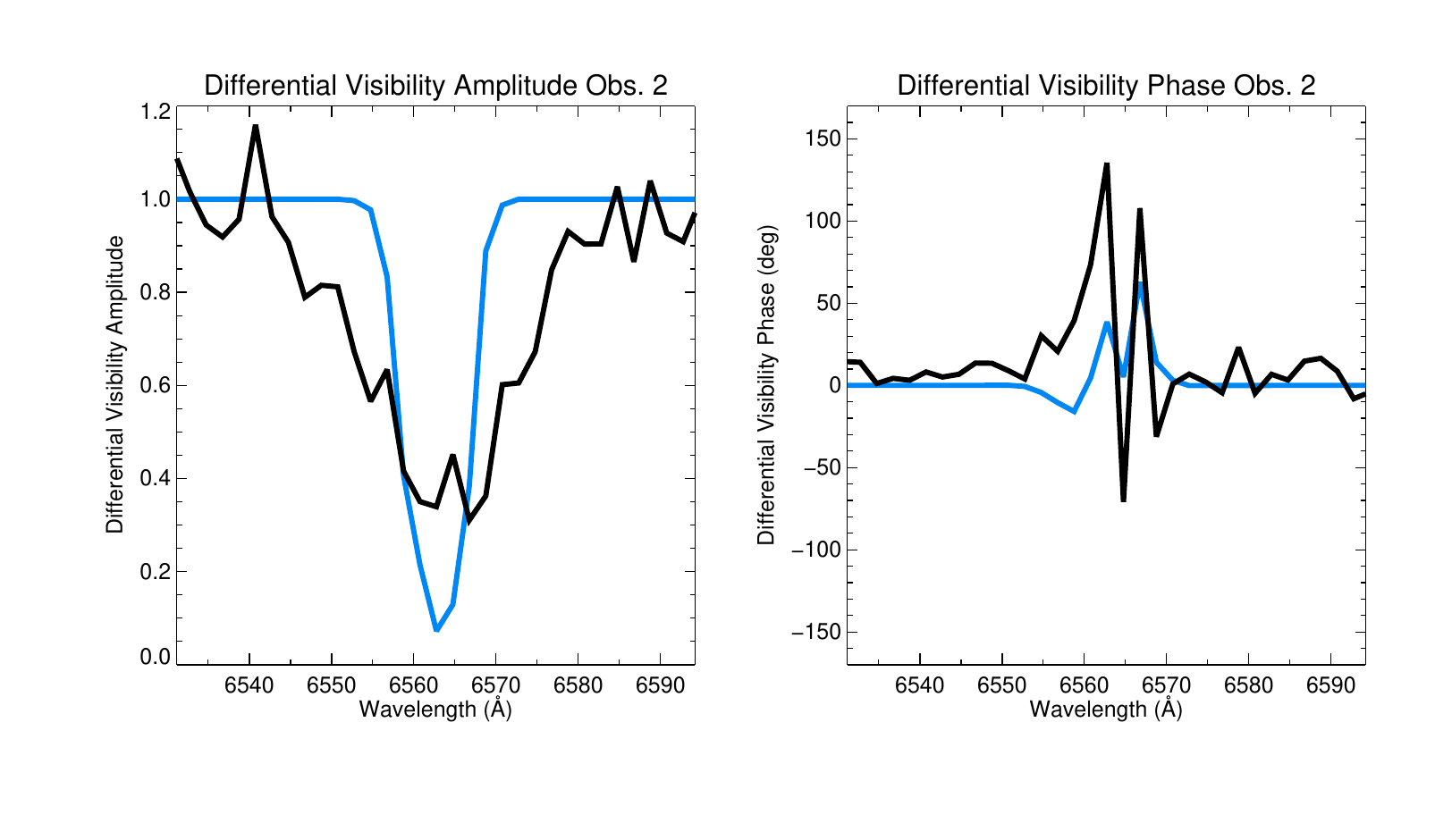}
}
\hspace{0mm}
\subfloat[2020 Dec 17, W2-W1, $(106.31,10.99)$]{
  \includegraphics[scale=0.3]{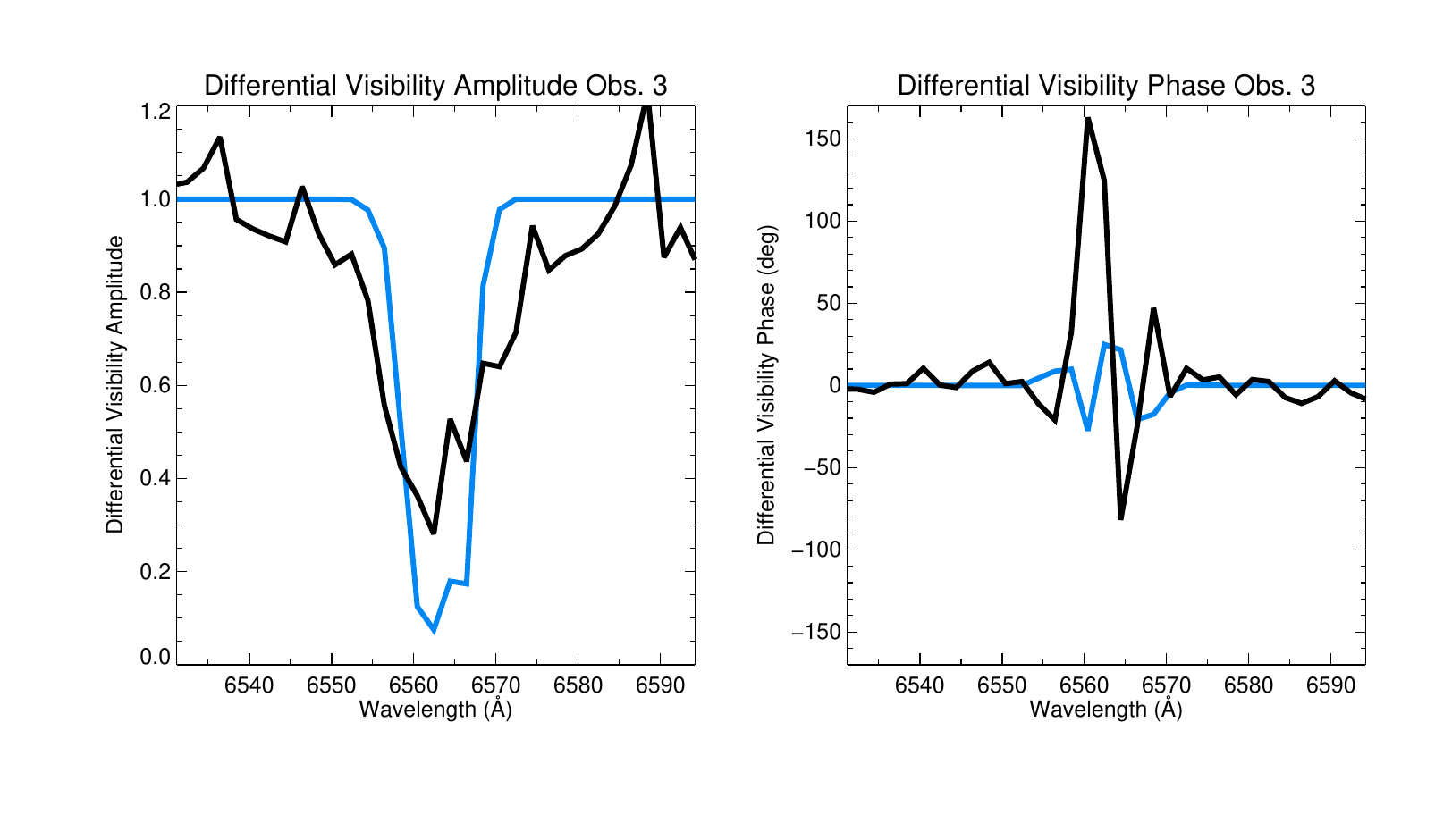}
}
\subfloat[2020 Dec 19, E2-E1, $(-58.69,-22.53)$]{
  \includegraphics[scale=0.3]{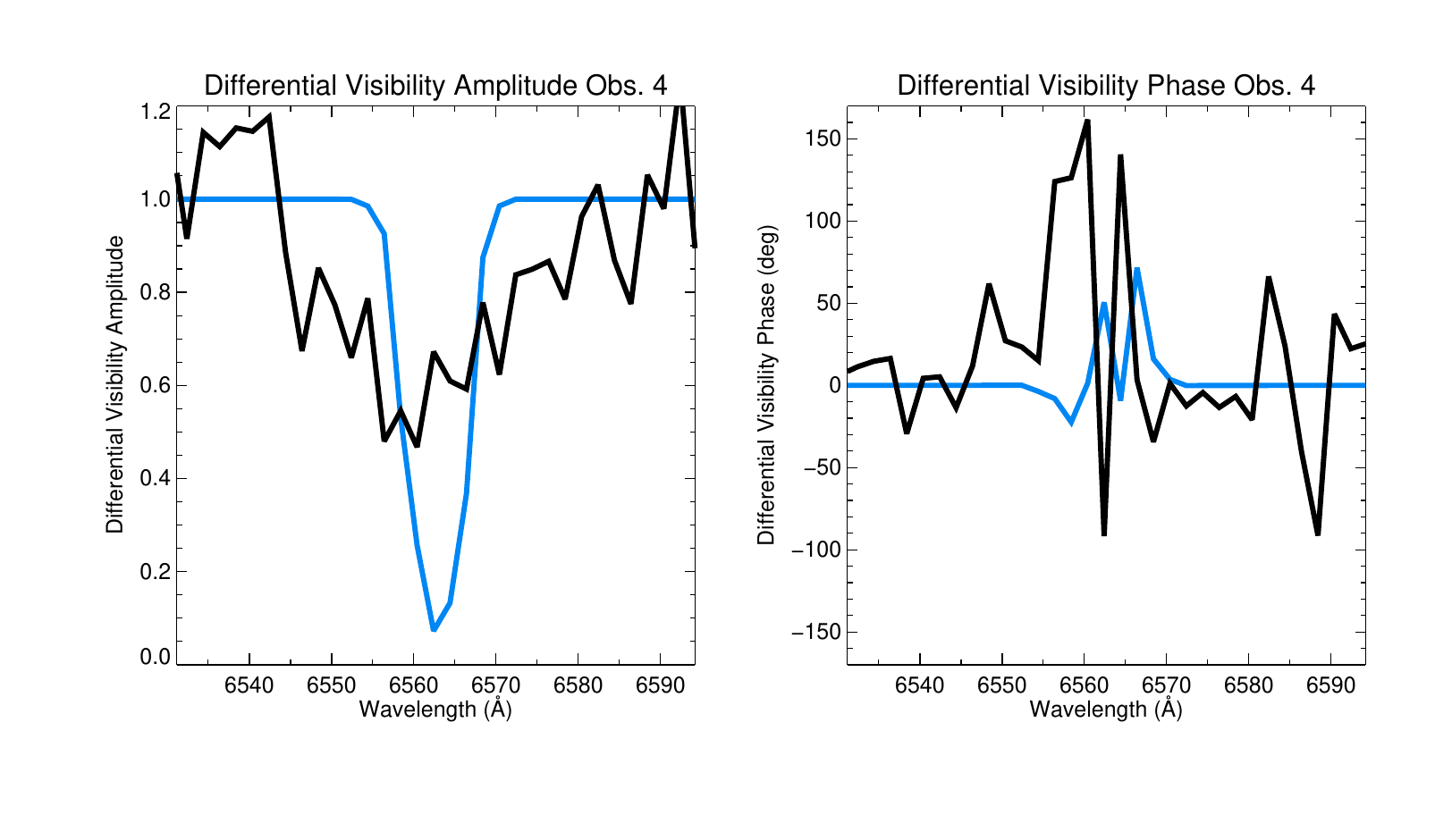}
}
\caption{Differential interferometric visibility amplitude and phase for $\gamma$ Cas.  Black lines show the observations 
and blue marks the best fit asymmetric model. Each plot is indicated by the date of observation, telescope pair used, 
and spatial frequency coordinates $(u,v)$.}
\end{figure}

\newpage

\FloatBarrier
\section{$\phi$ Per Interferometric Fits}

\begin{figure}[h] 
\centering
\subfloat[2020 Nov 23, S1-S2, $(18.56,-21.74)$]{
  \includegraphics[scale=0.3]{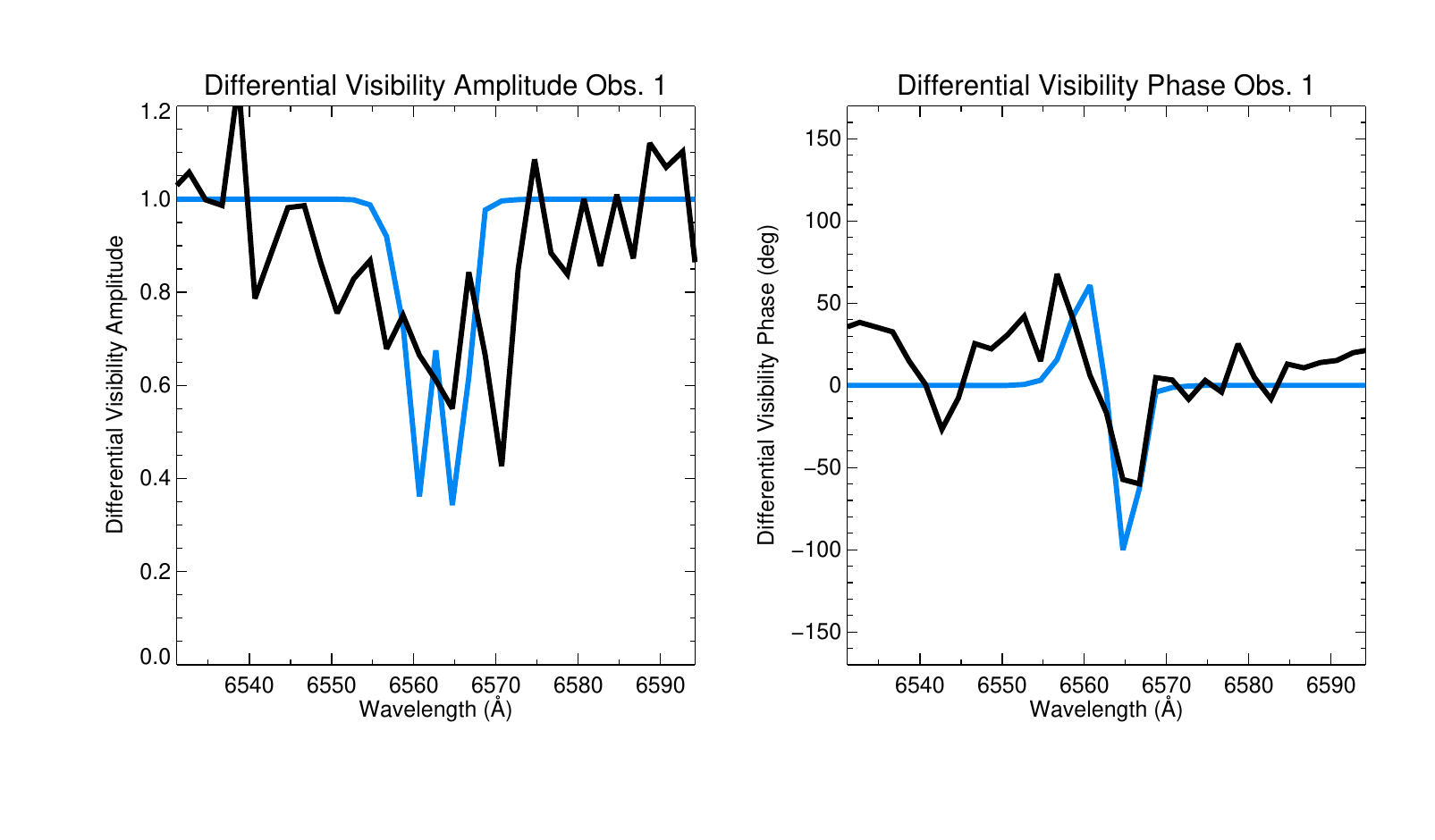} 
}
\subfloat[2020 Dec 16, W2-W1, $(43.80,71.19)$]{
  \includegraphics[scale=0.3]{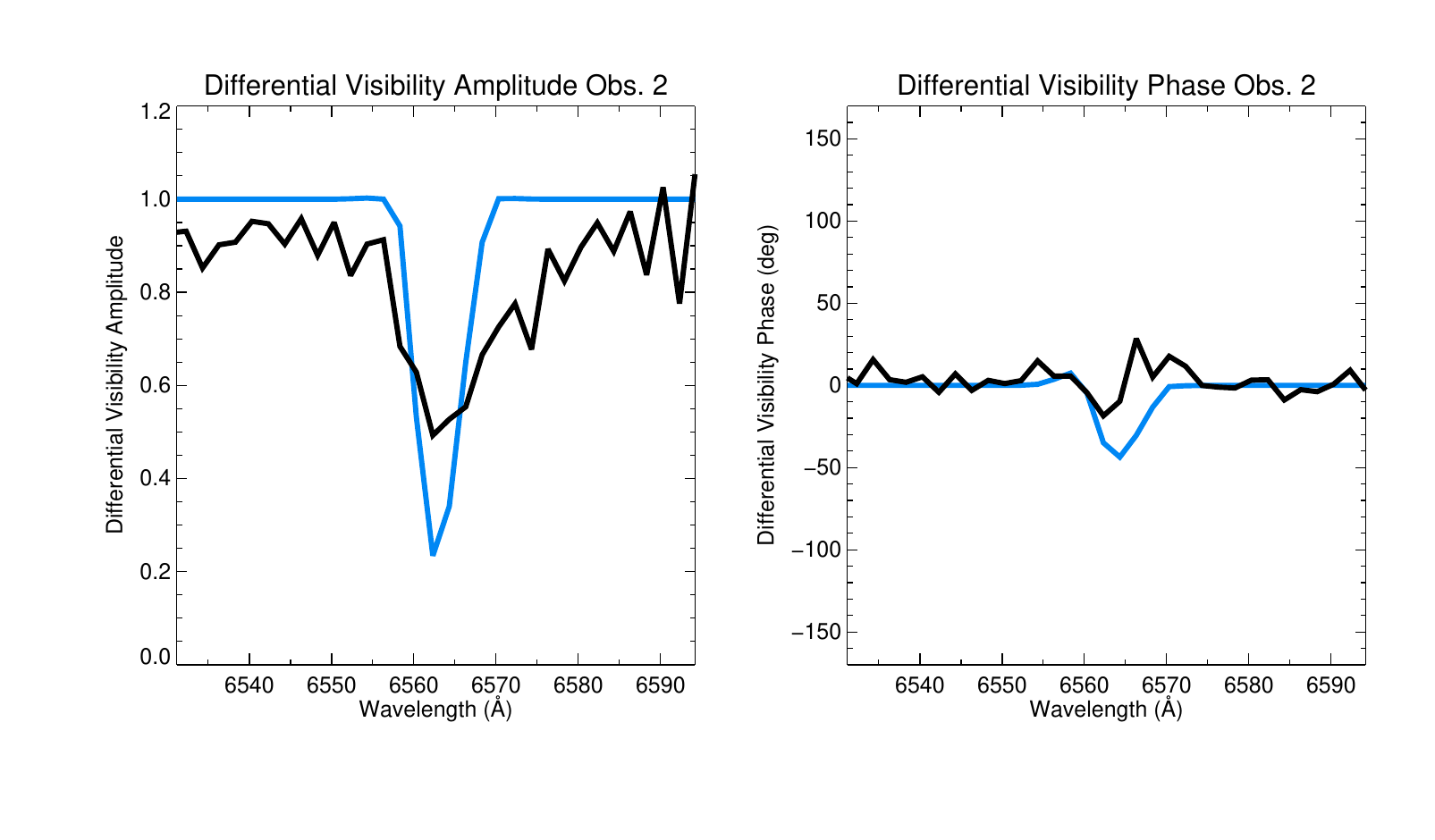}
}
\hspace{0mm}
\subfloat[2020 Dec 17, W2-W1, $(107.34,1.35)$]{
  \includegraphics[scale=0.3]{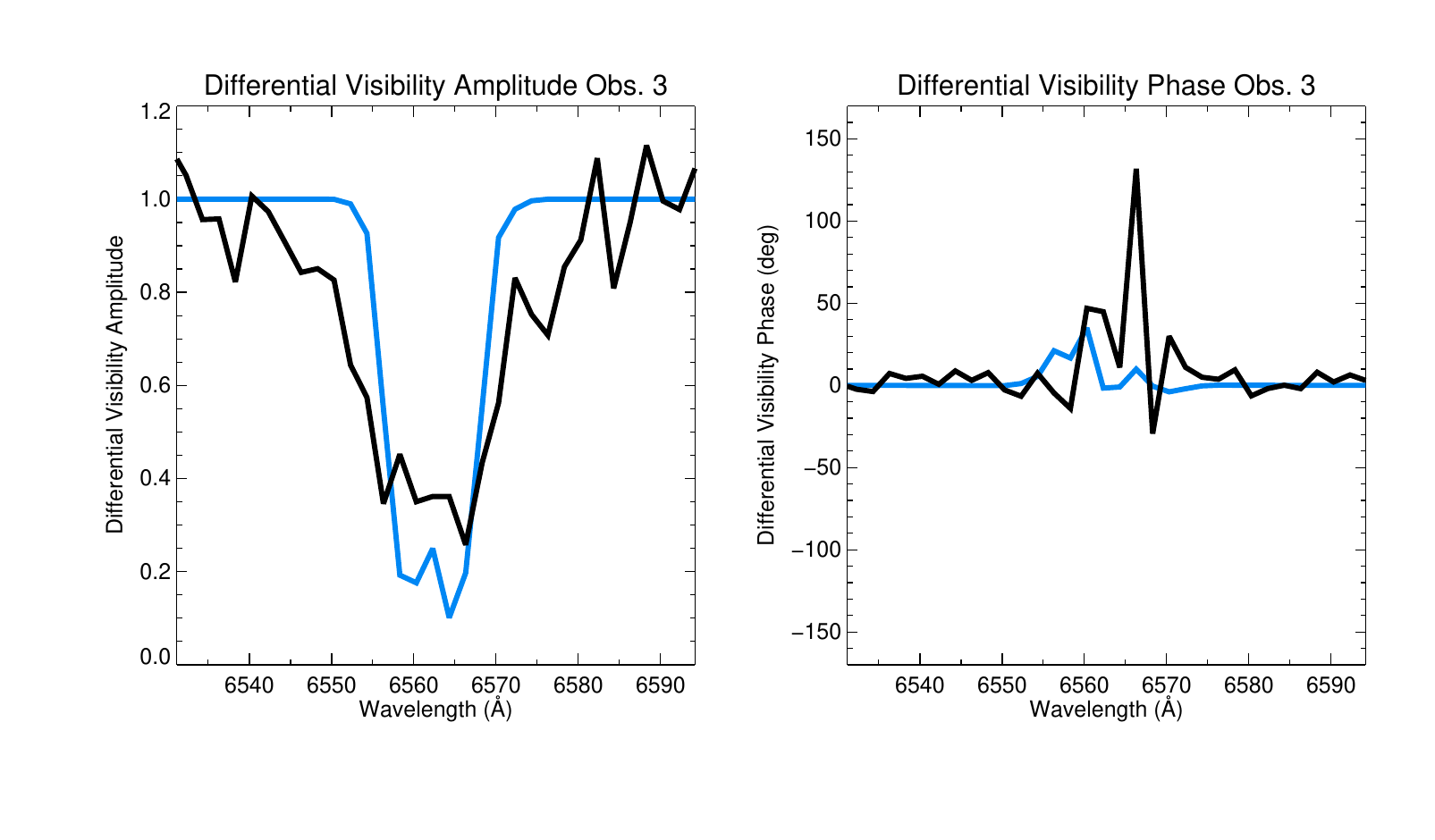}
}
\caption{Differential interferometric visibility amplitude and phase for $\phi$ Per in the same format as Figure 8.}
\end{figure}

\newpage

\FloatBarrier
\section{$\zeta$ Tau Interferometric Fits}

\begin{figure}[h] 
\centering
\subfloat[2020 Nov 21, S1-S2, $(18.61,-27.99)$]{
  \includegraphics[scale=0.27]{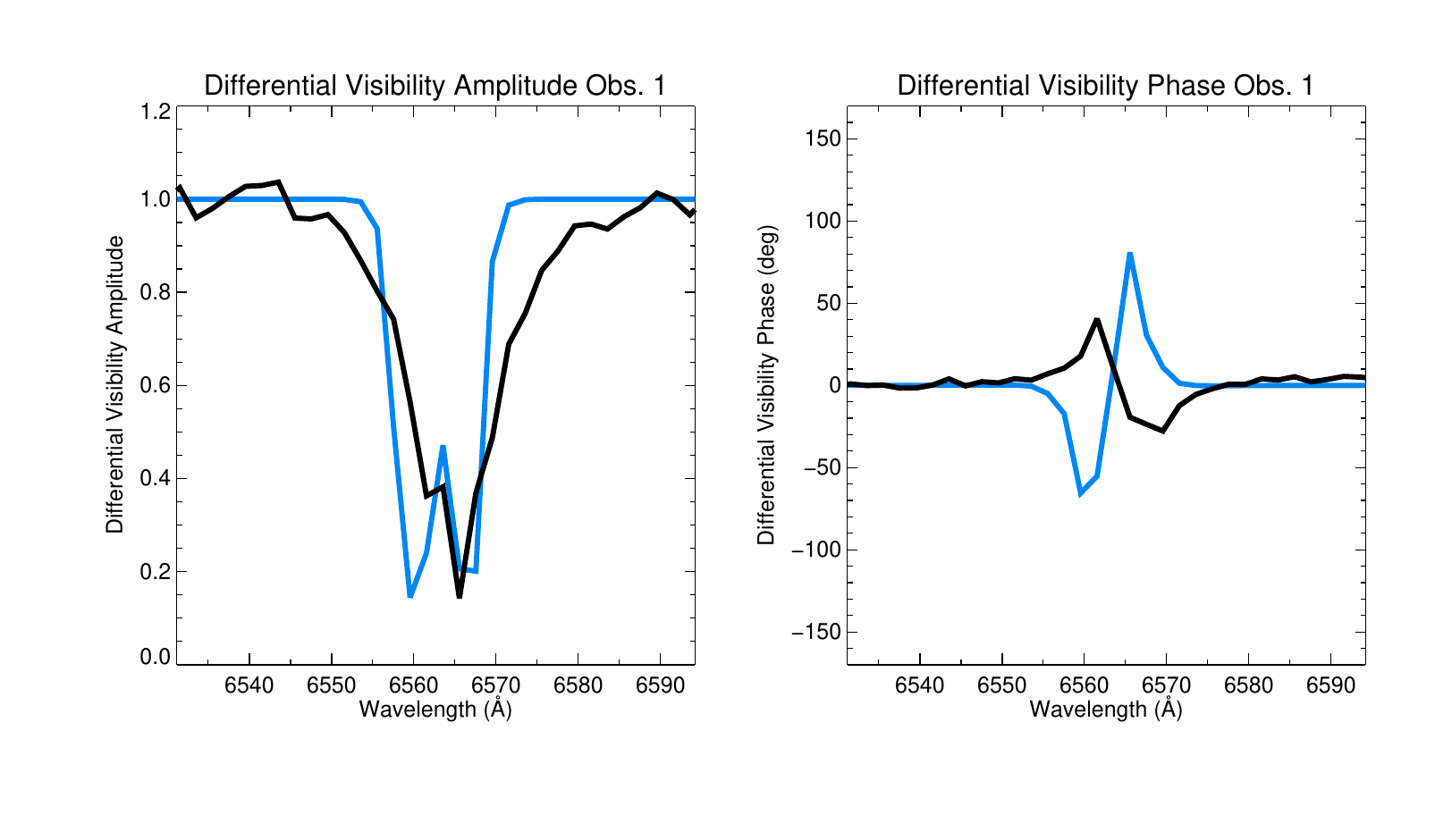}
}
\subfloat[2020 Nov 22, E2-E1, $(-12.78,-47.27)$]{
  \includegraphics[scale=0.27]{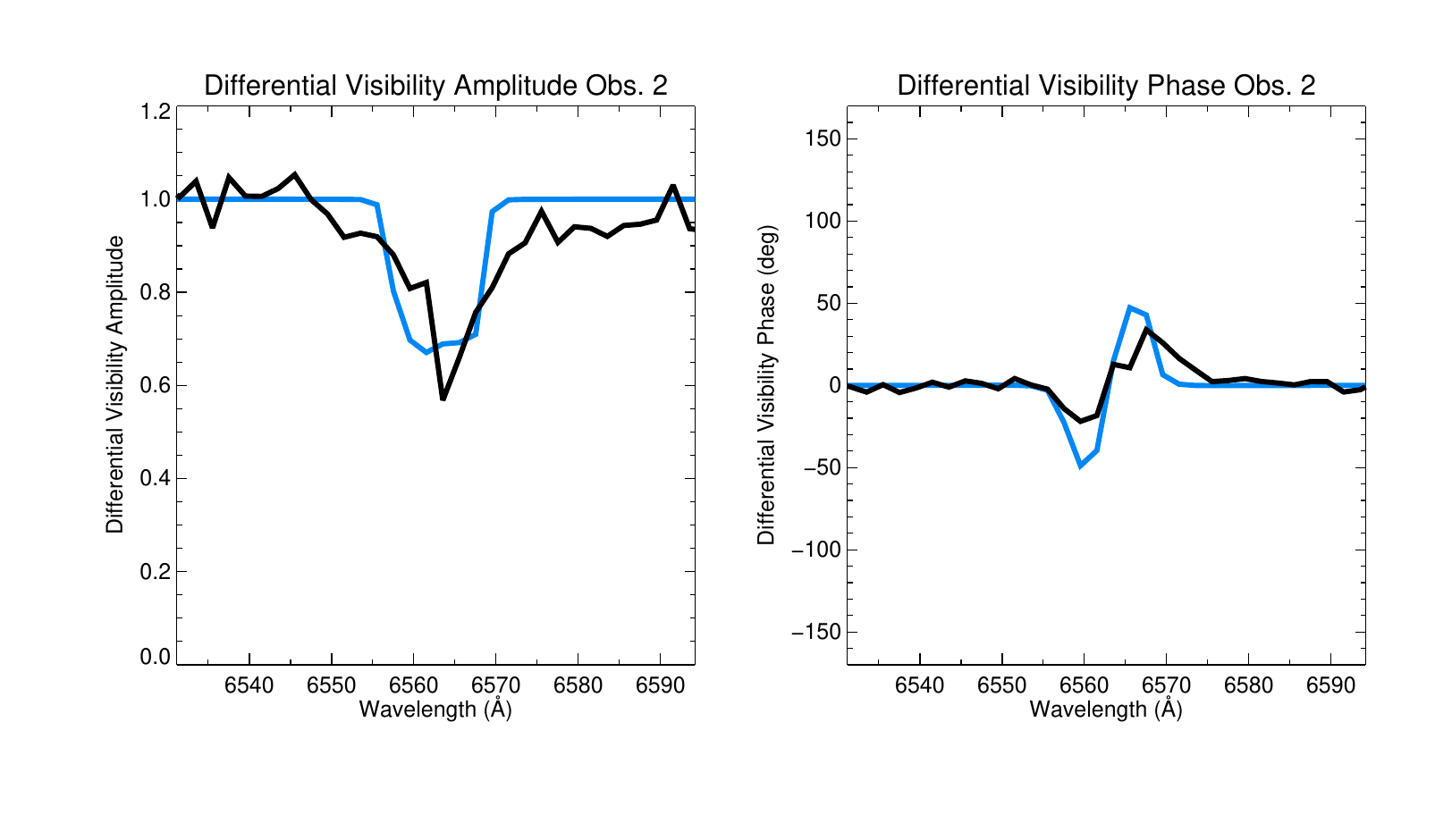}
}
\hspace{0mm}
\subfloat[2020 Nov 23, S1-S2, $(14.58,-30.76)$]{
  \includegraphics[scale=0.27]{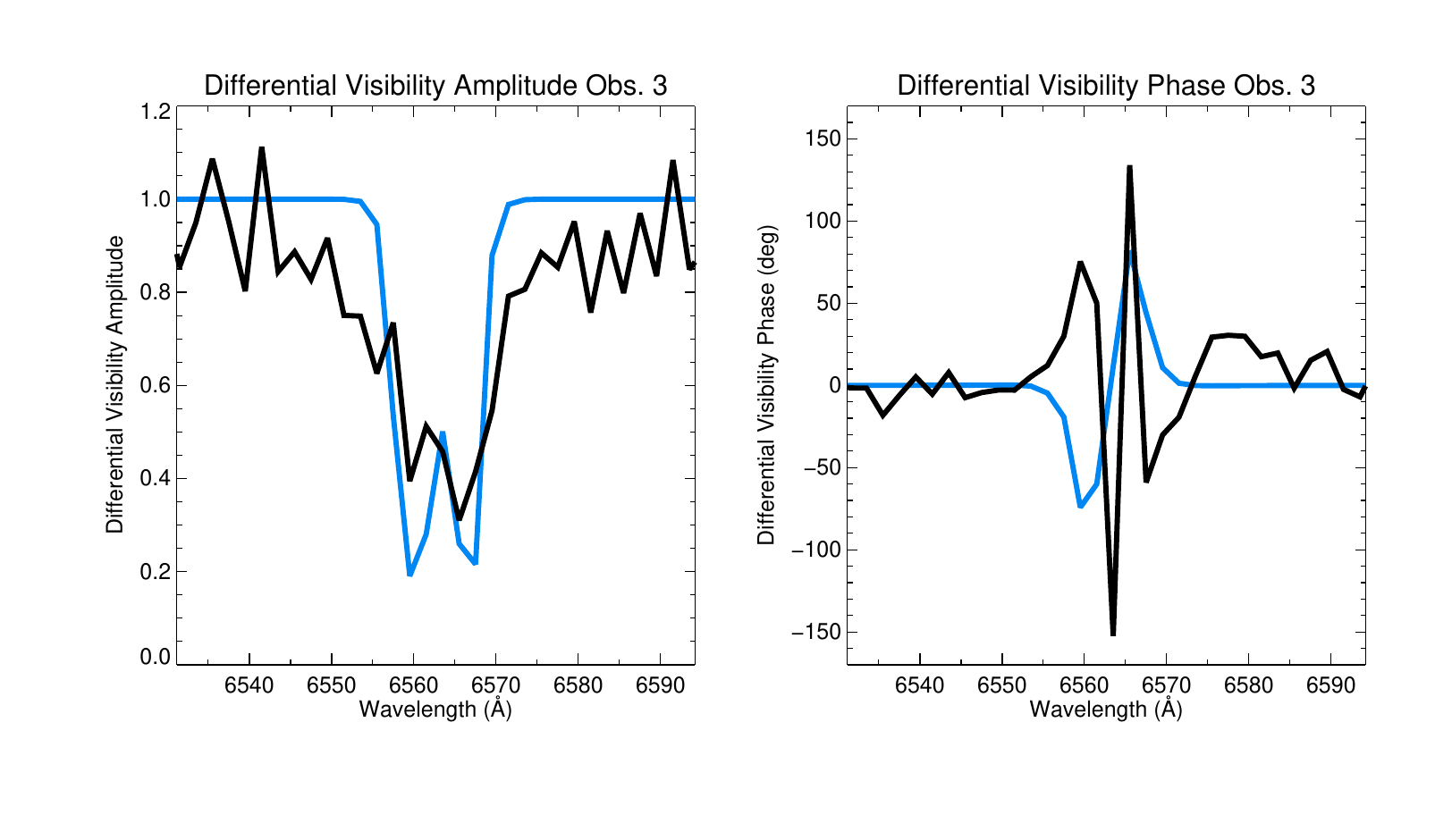}
}
\subfloat[2020 Nov 23, S1-S2, $(16.73,-29.66)$]{
  \includegraphics[scale=0.27]{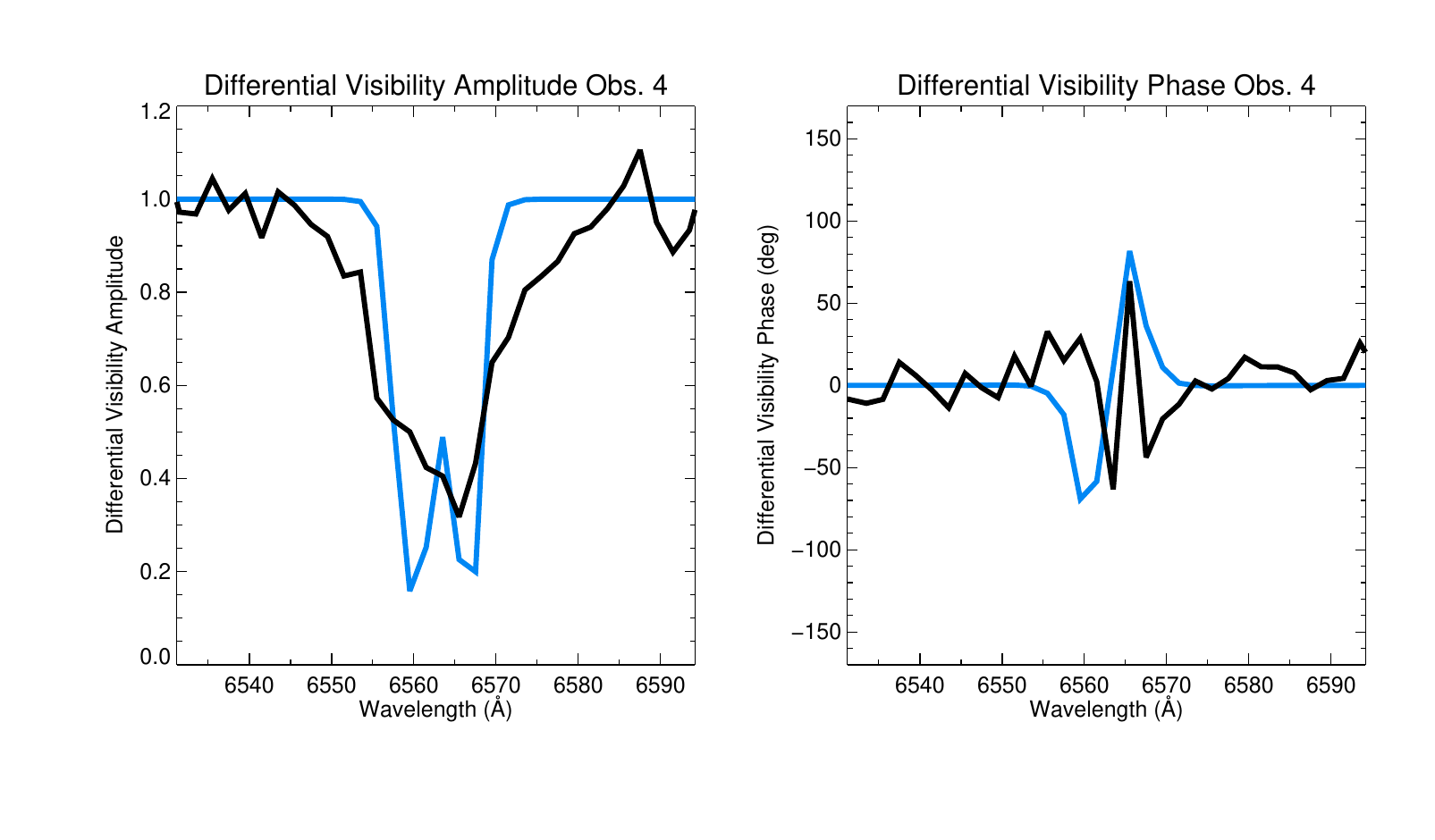}
}
\hspace{0mm}
\subfloat[2020 Dec 16, W2-W1, $(102.43,4.74)$]{
  \includegraphics[scale=0.27]{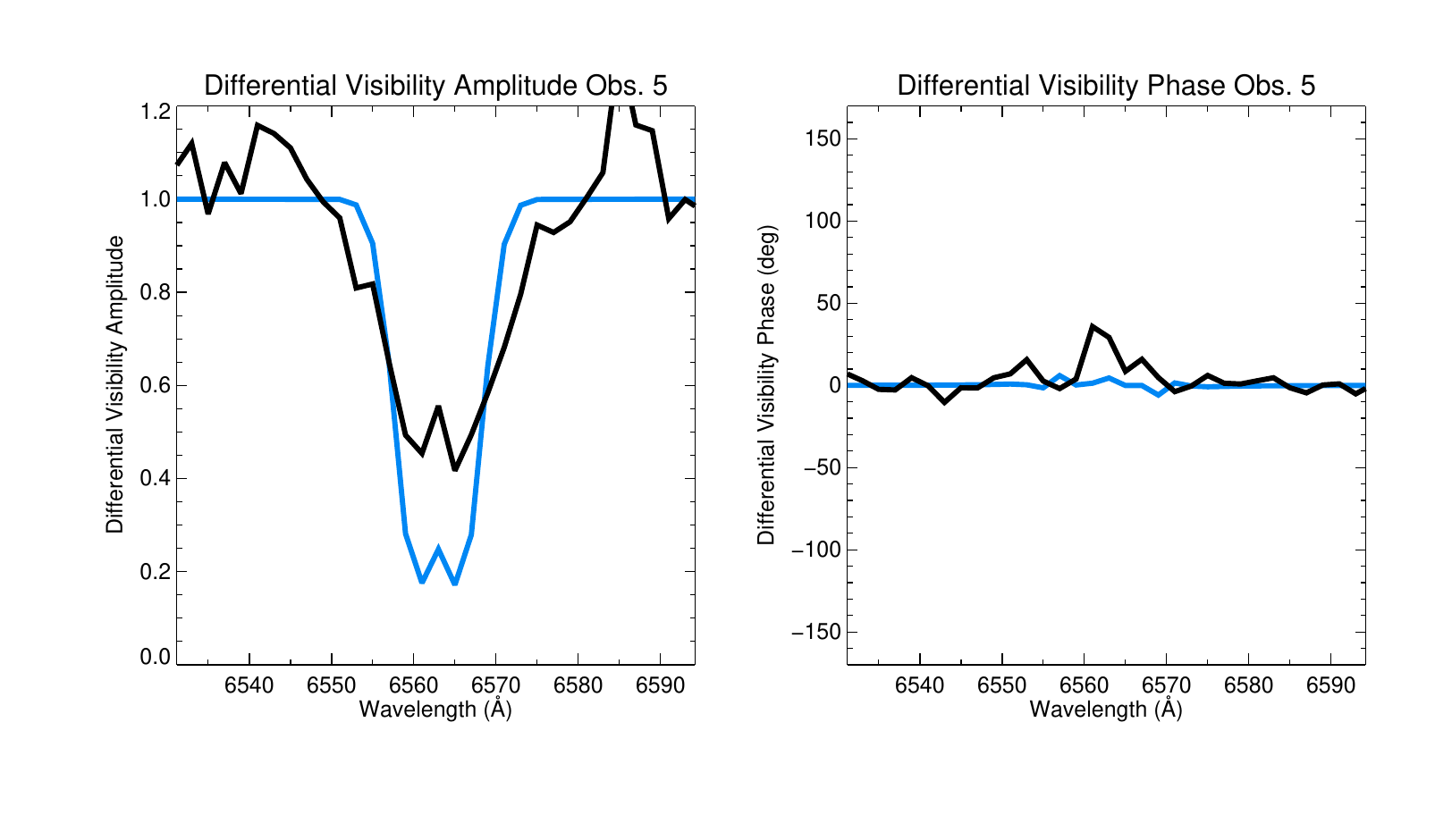}
}
\subfloat[2020 Dec 16, W2-W1, $(56.08,25.95)$]{
  \includegraphics[scale=0.27]{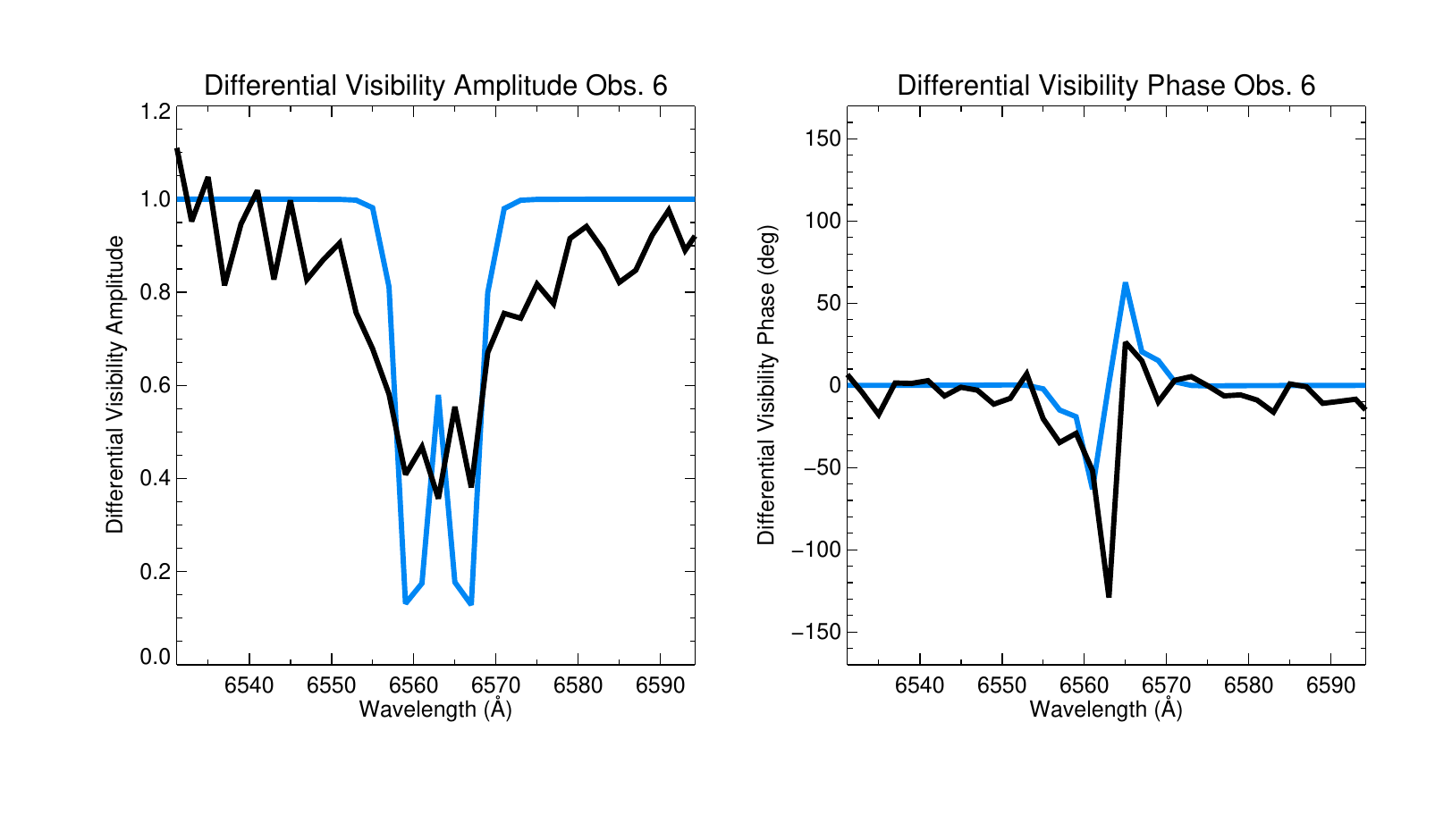}
}      
\hspace{0mm}
\subfloat[2020 Dec 17, W2-W1, $(80.12,-33.12)$]{
  \includegraphics[scale=0.27]{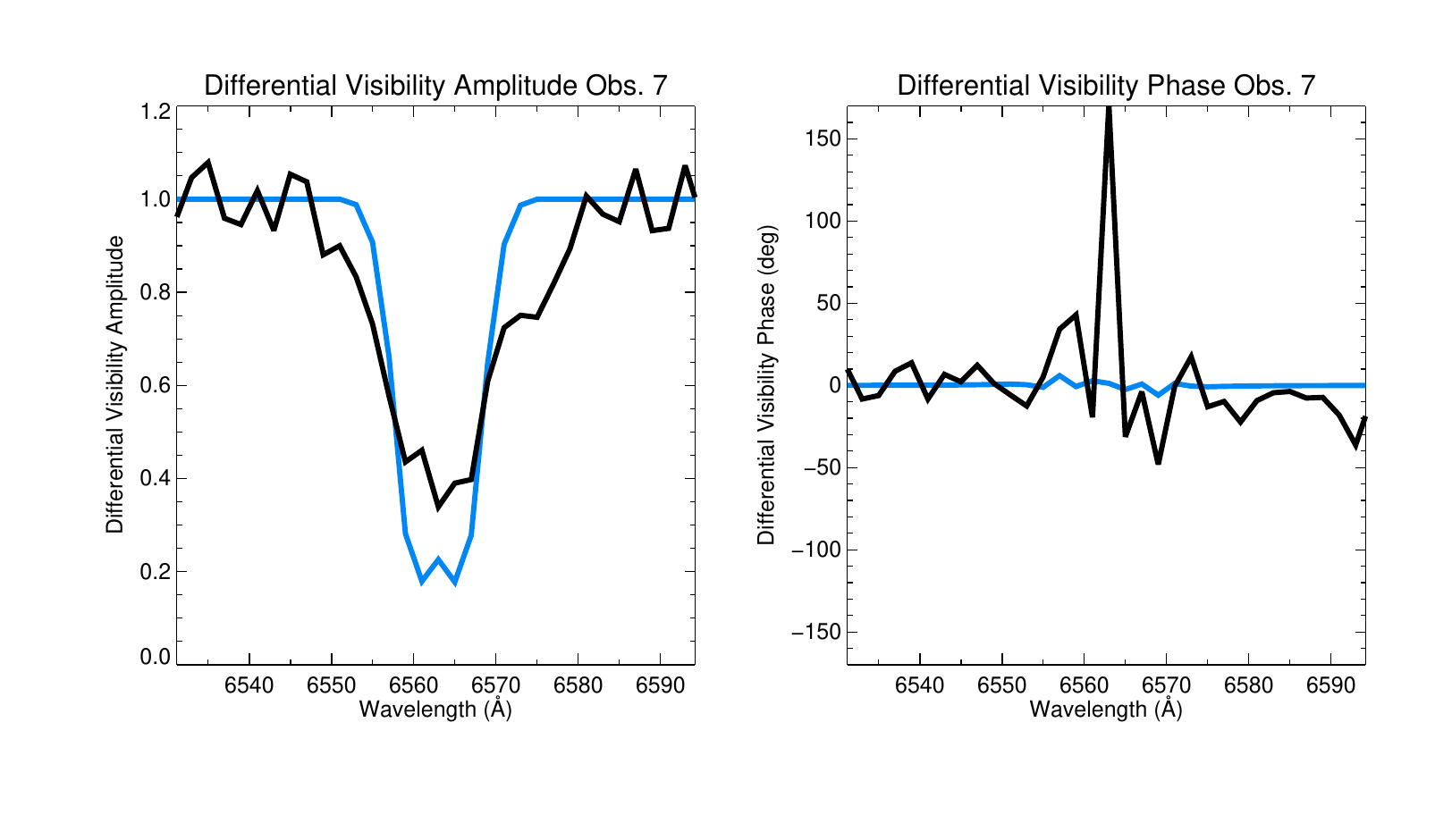}
}
\caption{Differential interferometric visibility amplitude and phase for $\zeta$ Tau in the same format as Figure 8.}
\end{figure}

\newpage

\FloatBarrier
\section{$\beta$ CMi Interferometric Fits}

\begin{figure}[h] 
\centering
\subfloat[2020 Nov 21, S1-S2, $(0.38,-30.60)$]{
  \includegraphics[scale=0.21]{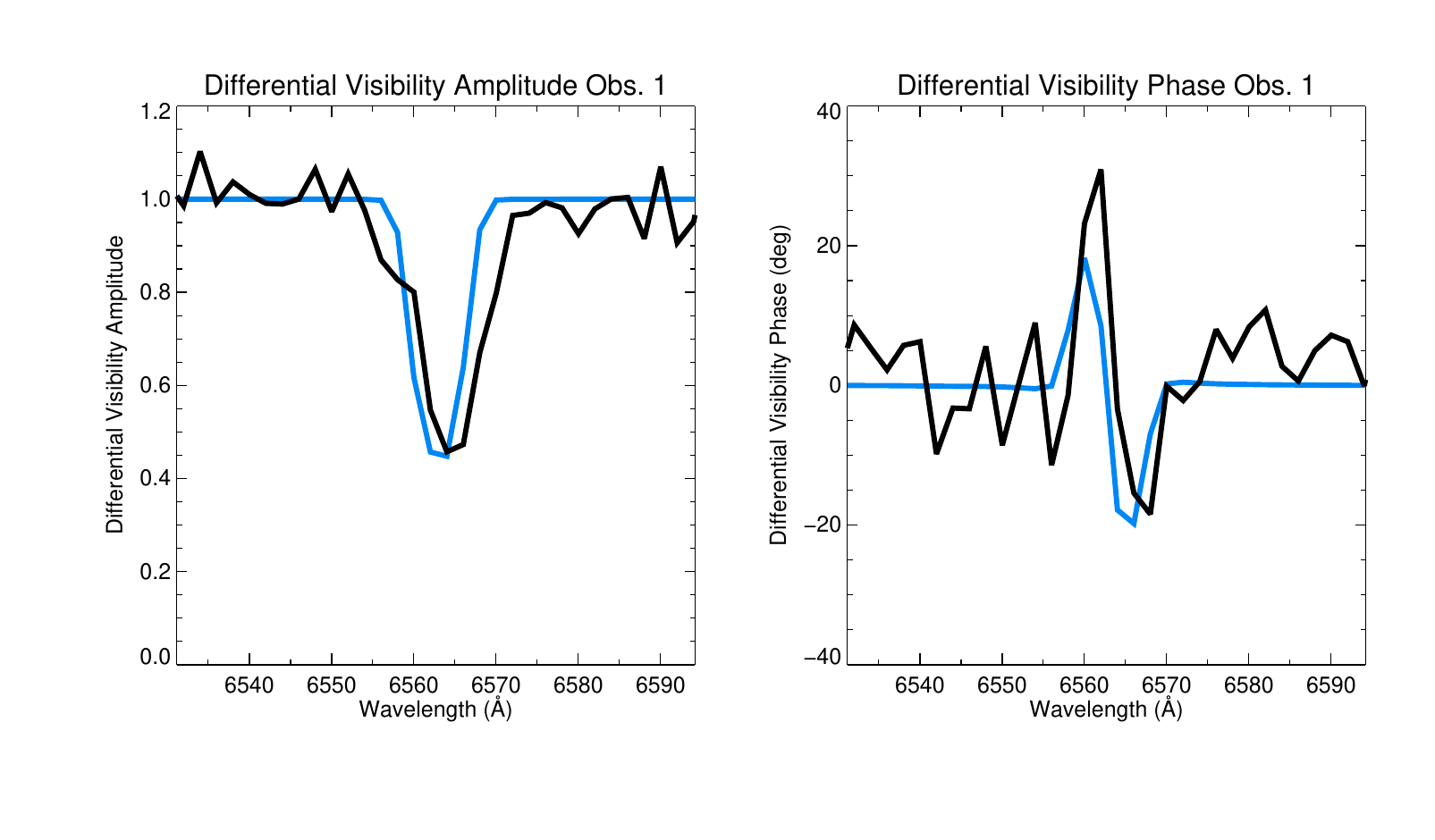}
}
\subfloat[2020 Nov 21, E2-E1, $(-48.75,-32.83)$]{
  \includegraphics[scale=0.21]{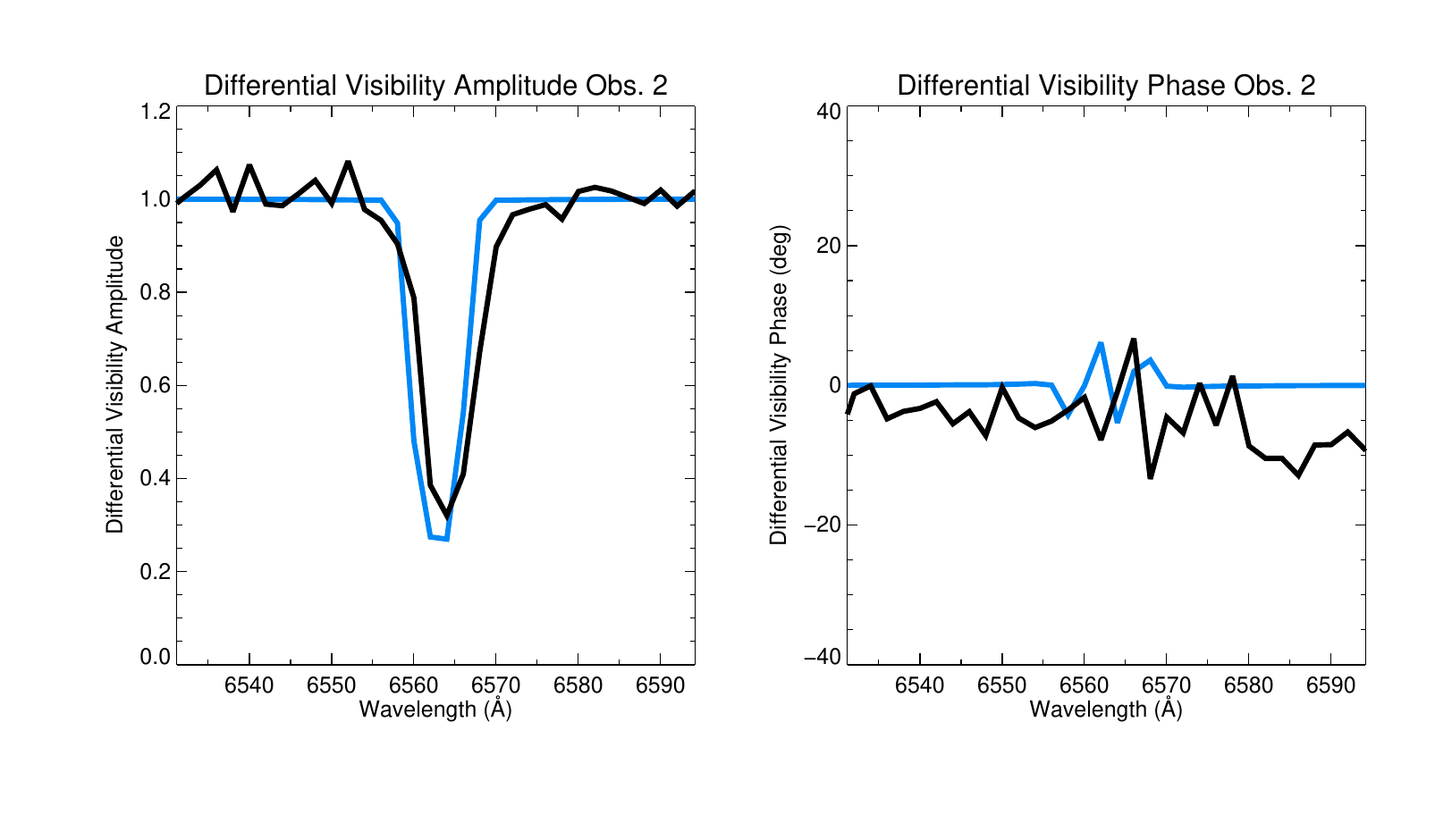}
}
\hspace{0mm}
\subfloat[2020 Nov 21, S1-S2, $(15.79,-29.42)$]{
  \includegraphics[scale=0.21]{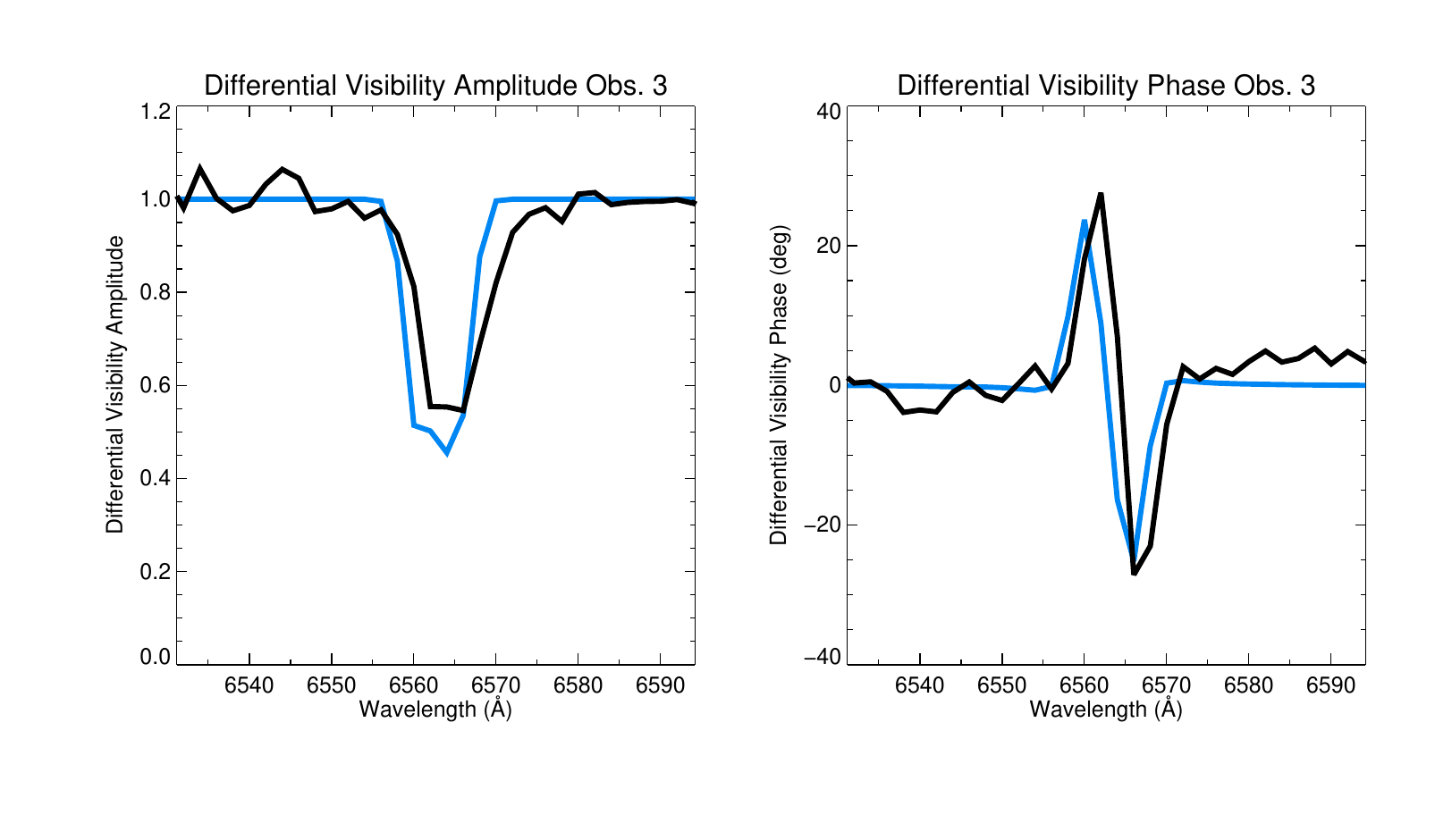}
}
\subfloat[2020 Nov 22, E2-E1, $(-32.87,-35.06)$]{
  \includegraphics[scale=0.21]{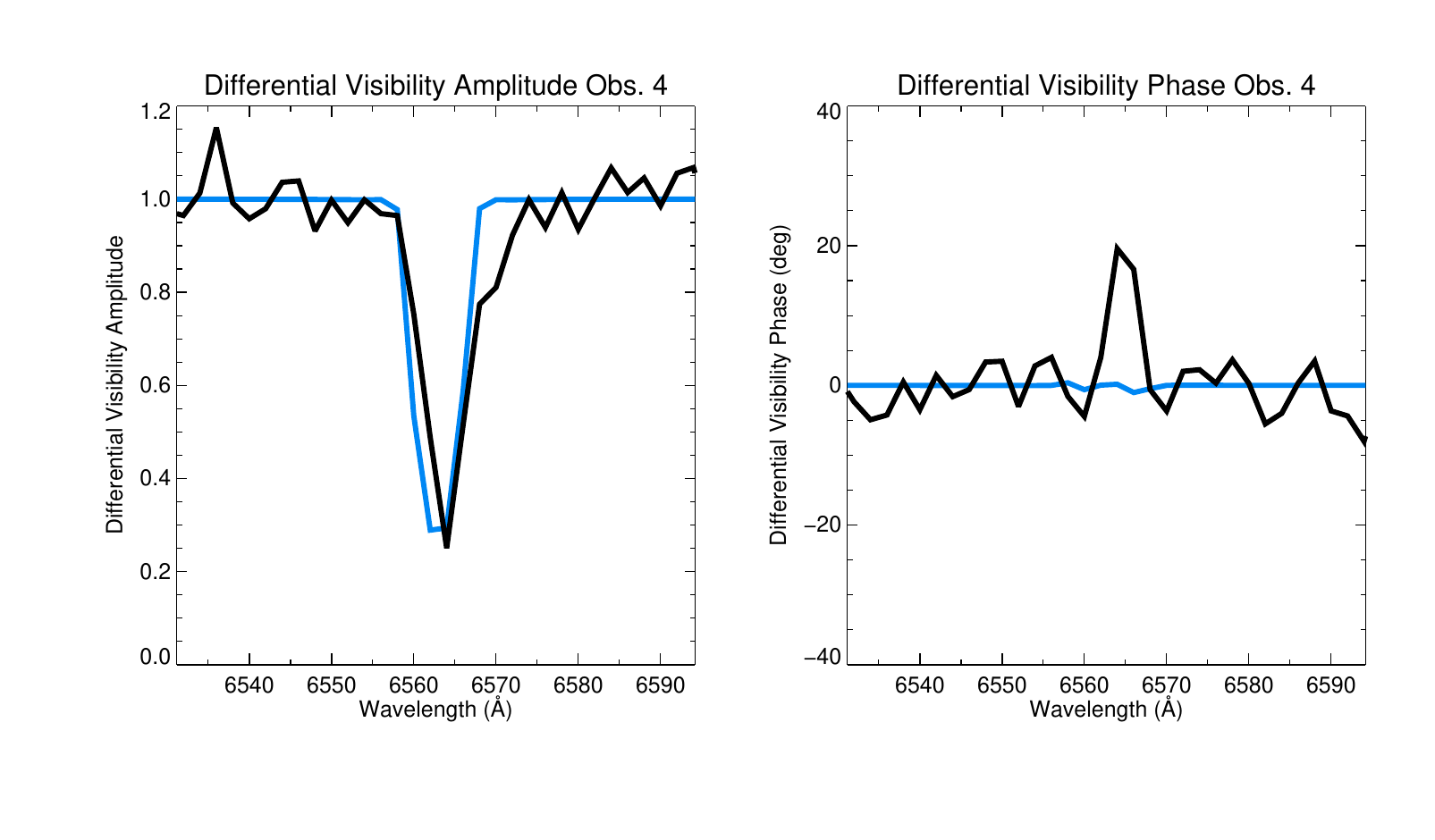}
}
\hspace{0mm}
\subfloat[2020 Nov 23, S1-S2, $(9.04,-30.28)$]{
  \includegraphics[scale=0.21]{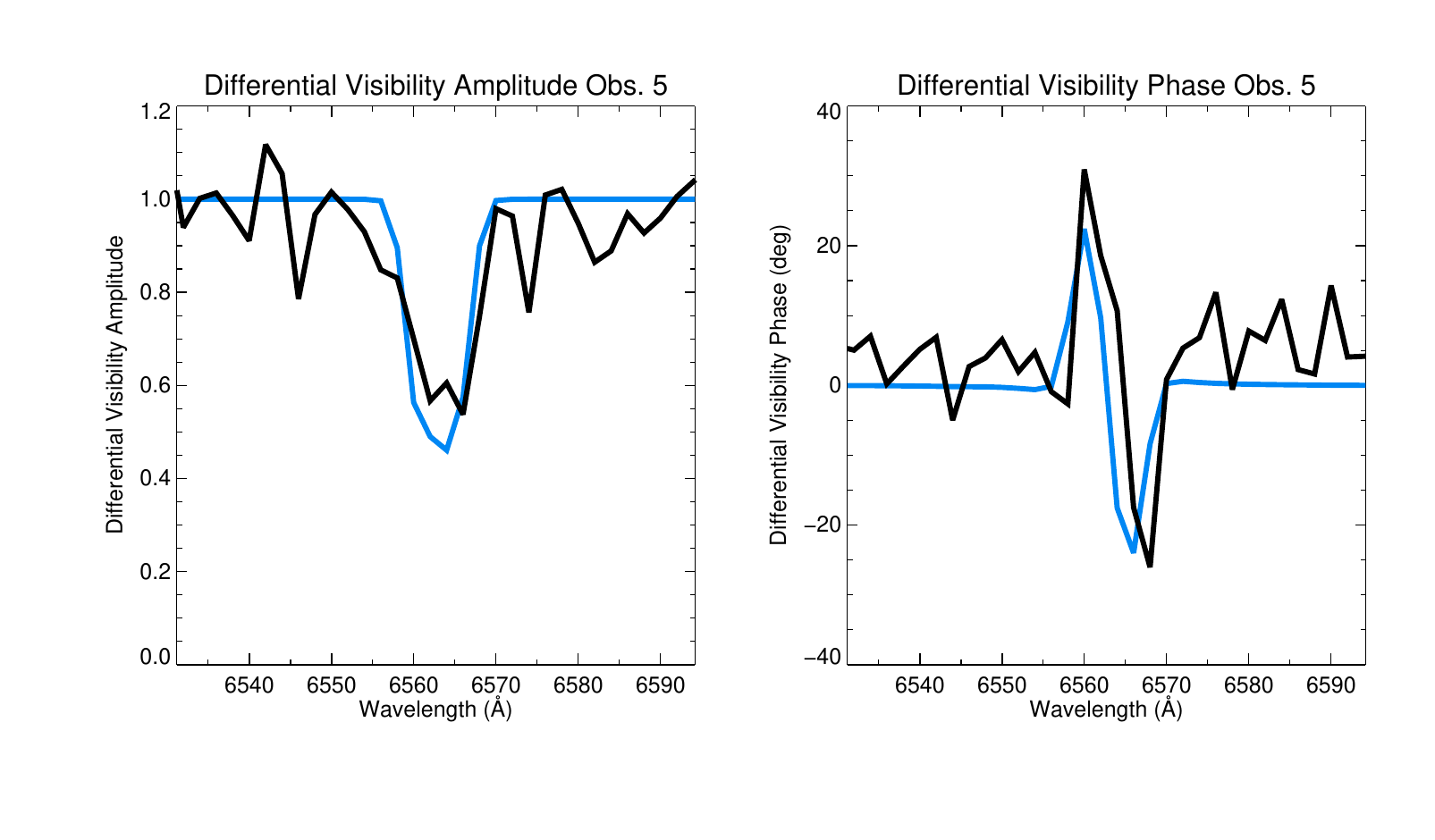}
}
\subfloat[2020 Nov 24, E2-E1, $(-50.66,-32.41)$]{
  \includegraphics[scale=0.21]{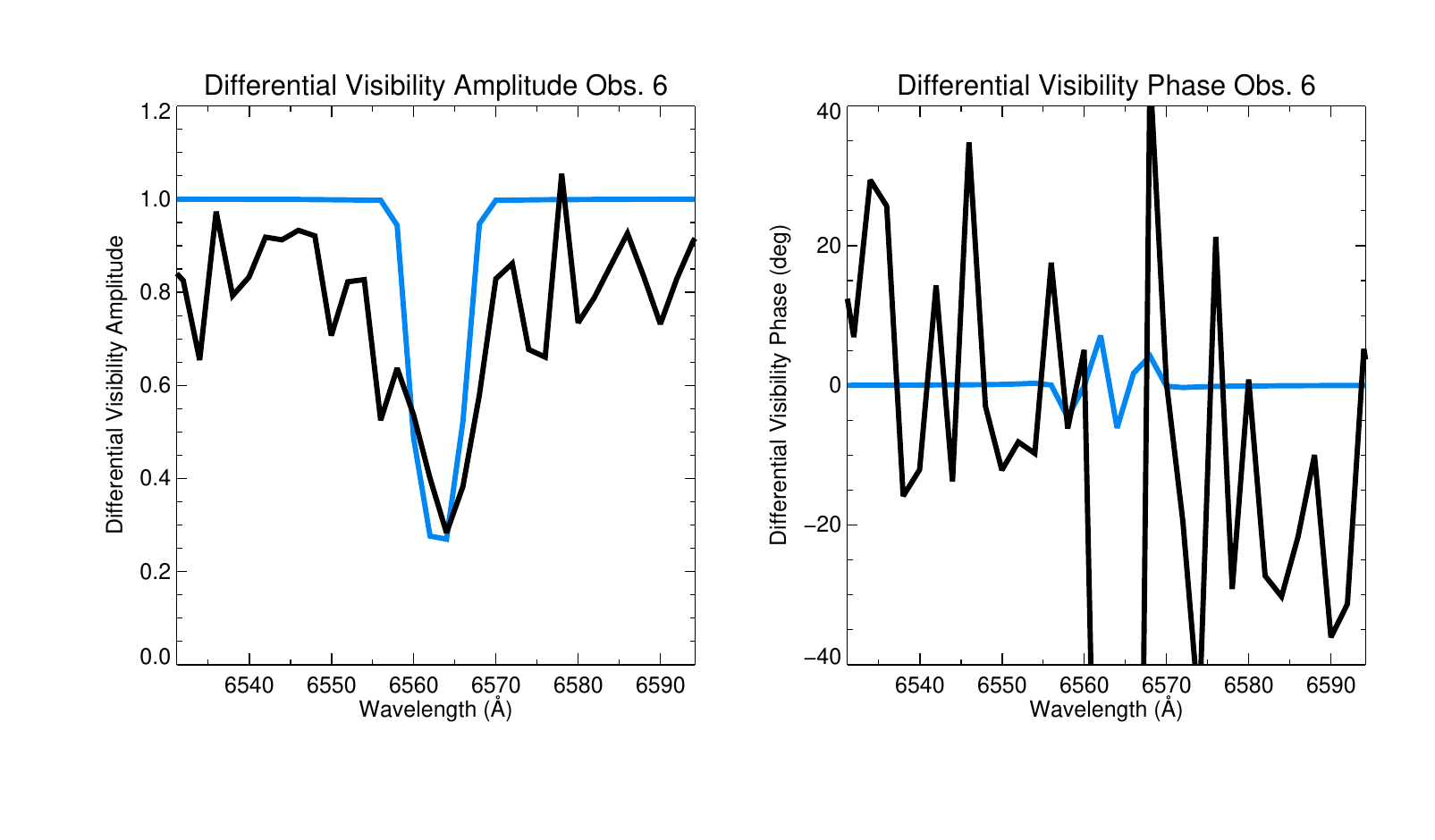}
}
\hspace{0mm}
\subfloat[2020 Nov 24, E2-E1, $(-44.28,-33.64)$]{
  \includegraphics[scale=0.21]{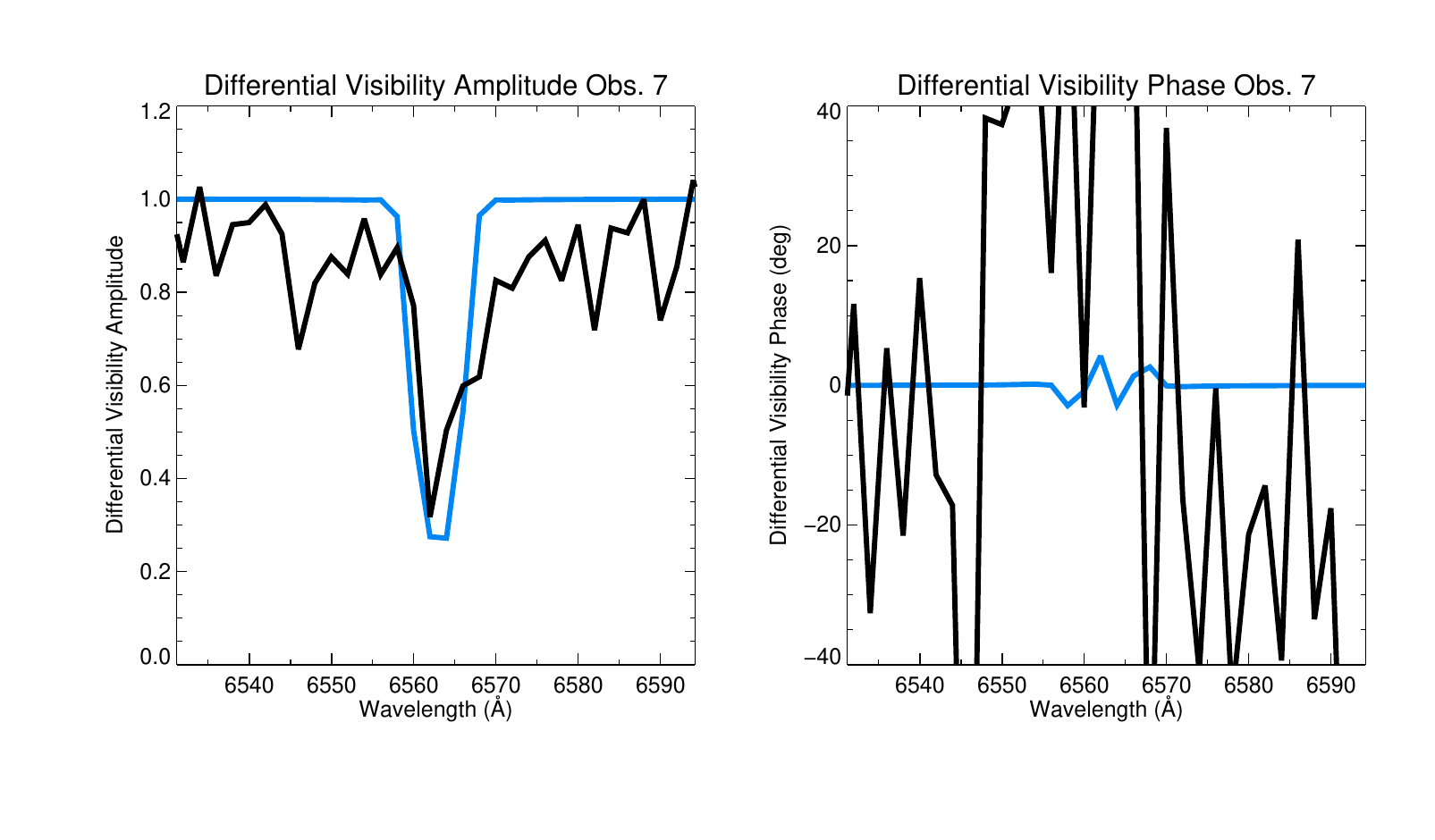}
}
\subfloat[2020 Nov 24, E2-E1, $(-35.42,-34.80)$]{
  \includegraphics[scale=0.21]{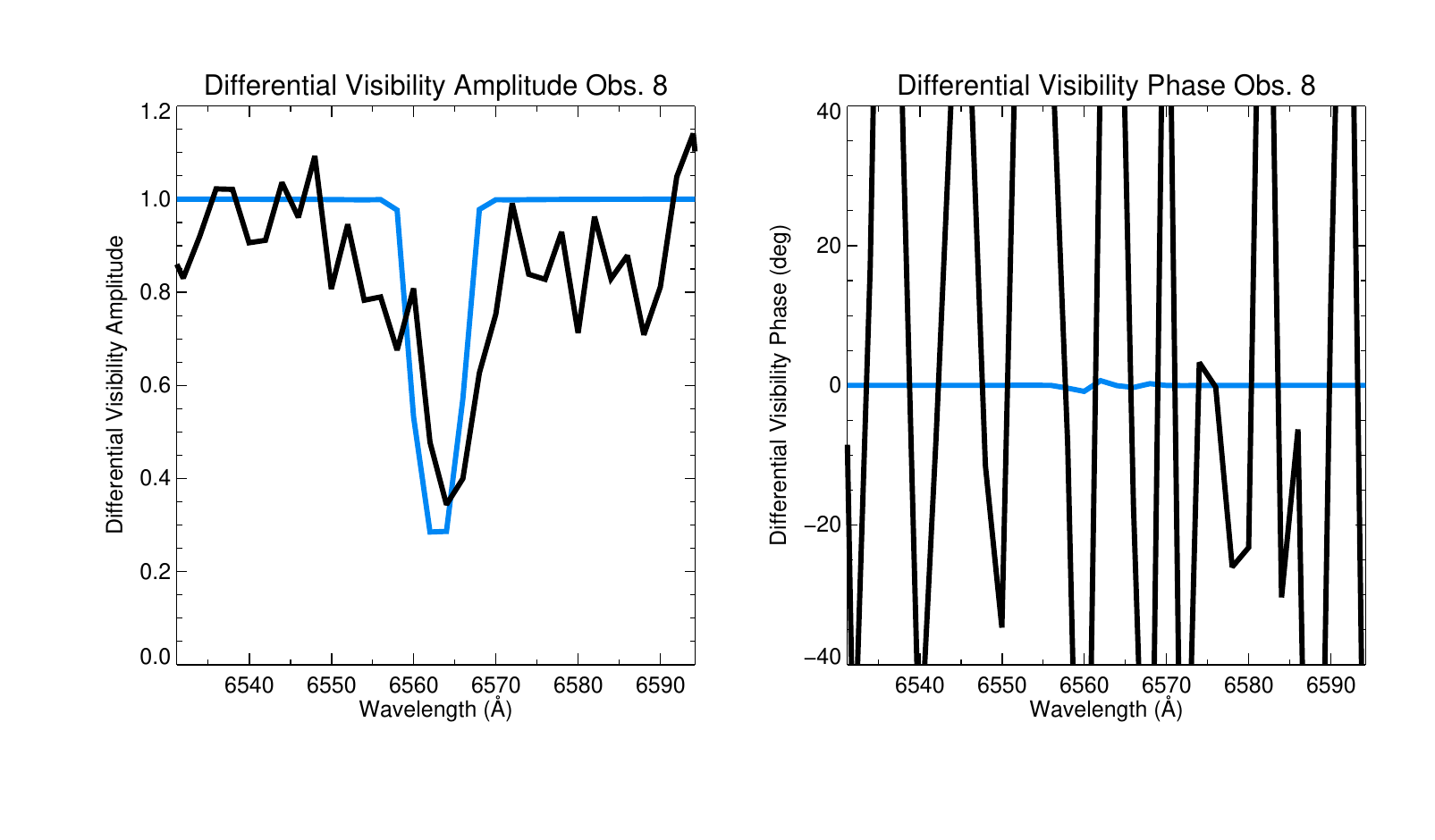}
}
\hspace{0mm}
\subfloat[2020 Dec 16, W2-W1, $(107.56,-8.26)$]{
  \includegraphics[scale=0.21]{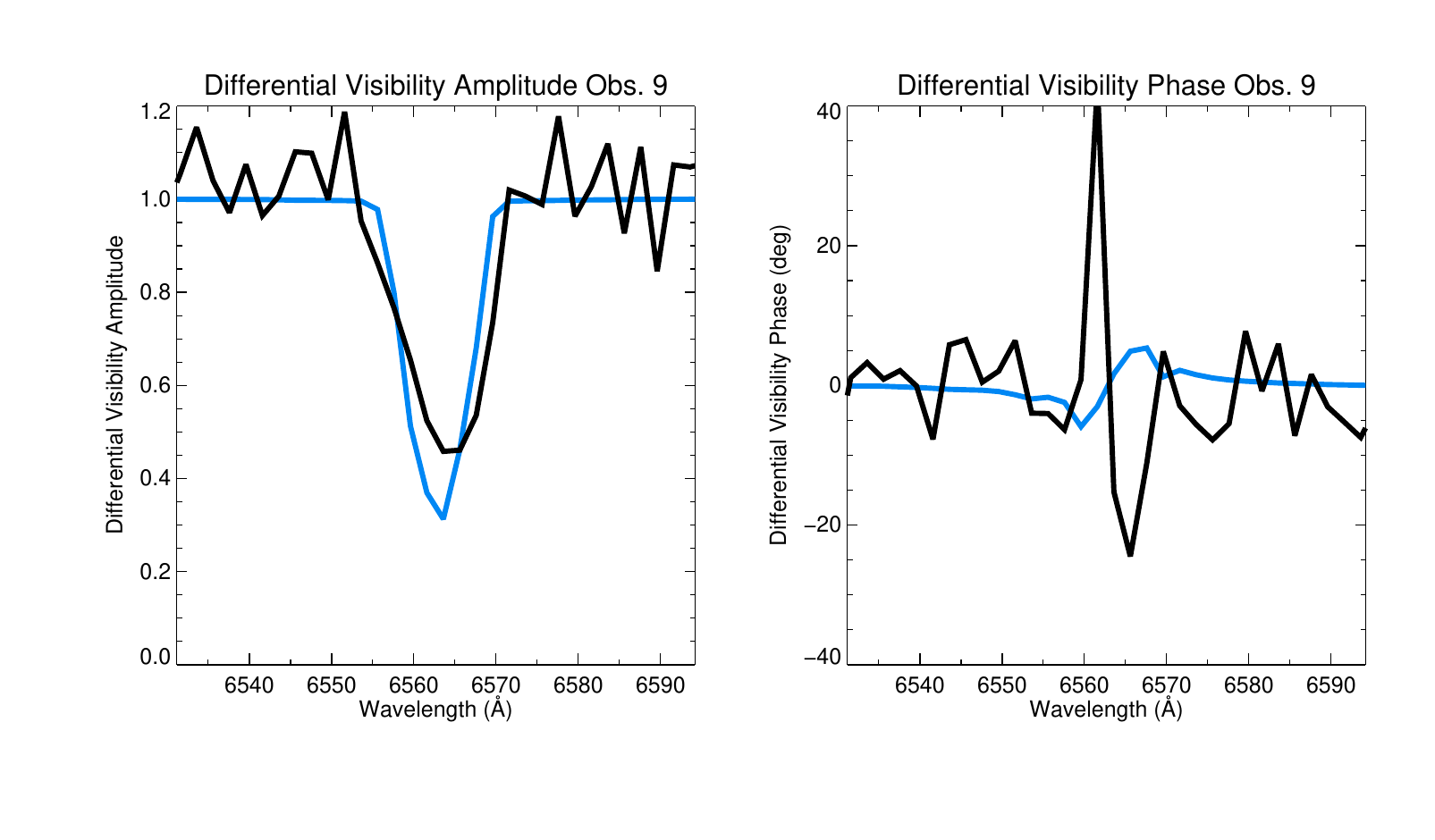}
}
\subfloat[2020 Dec 16, W2-W1, $(85.43,1.81)$]{
  \includegraphics[scale=0.21]{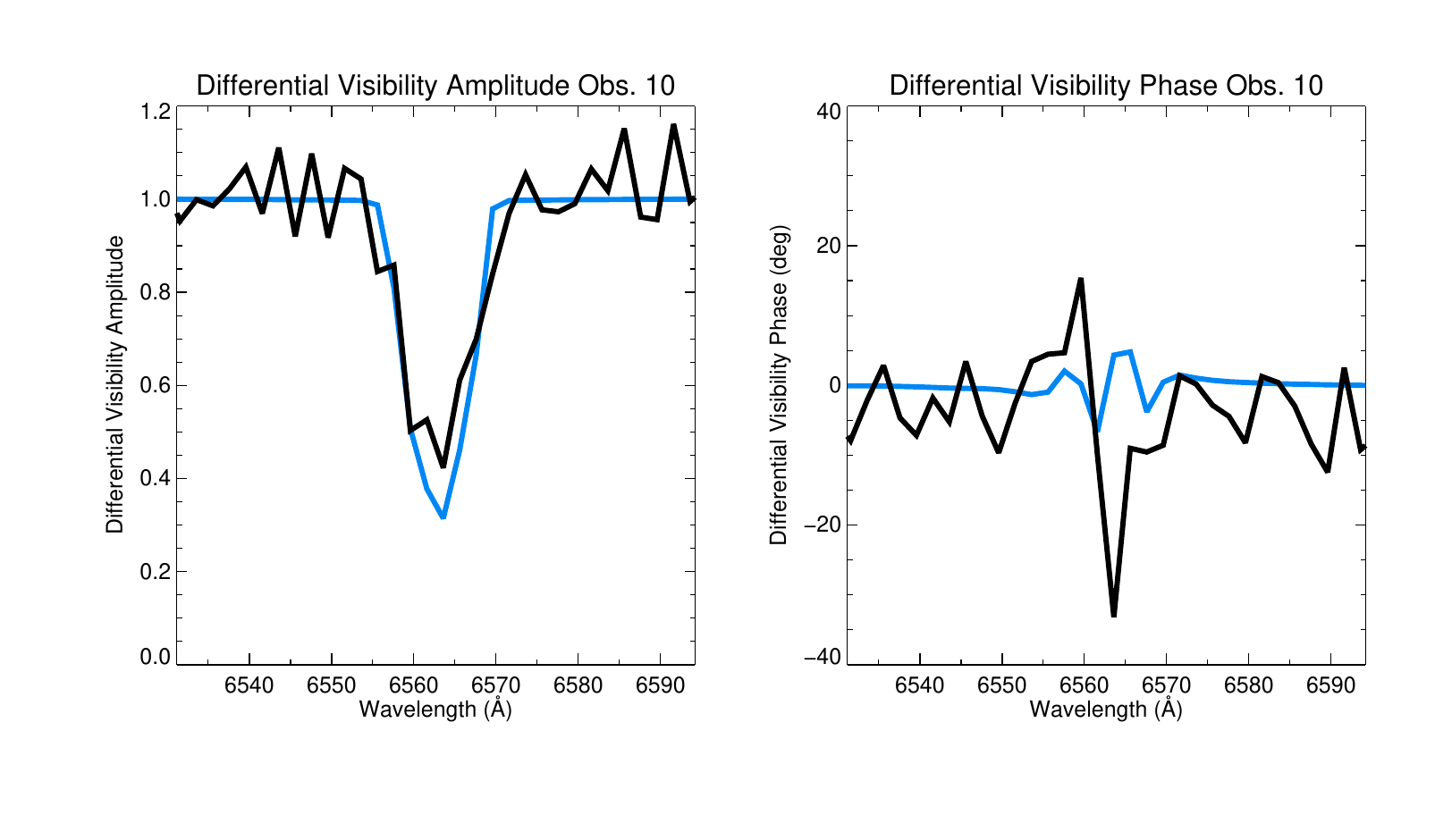}
}
\caption{Differential interferometric visibility amplitude and phase for $\beta$ CMi in the same format as Figure 8.}
\end{figure}

\newpage

\begin{acknowledgments}
We are grateful for insightful comments from an anonymous referee that 
led to many improvements in this paper.
This work is based upon observations obtained with the Georgia State University Center
for High Angular Resolution Astronomy Array at Mount Wilson Observatory. The CHARA
Array is supported by the National Science Foundation under Grant No. AST-2034336 and
AST-2407956. Institutional support has been provided from the GSU College of Arts and Sciences, 
Office of the Provost, and Office of the Vice President for Research and Economic Development.
VEGA was a collaboration between CHARA and Laboratoire
Fizeau (OCA/UNS/CNRS-Nice), LAOG in Grenoble, CRAL in Lyon, and
LESIA in Paris-Meudon. It was supported by French programs for stellar
physics and high angular resolution PNPS and ASHRA, by INSU-CNRS, and by
the R\'{e}gion PACA. 
This work has made use of the BeSS database, maintained at the Observatoire de
Meudon, France: http://basebe.obspm.fr.  We thank all the contributors to the BeSS,
especially Etienne Bertrand, Christian Buil, Albert Stiewing, and Olivier Thizy.
\end{acknowledgments}

\begin{contribution}
This work is based upon the Ph.D.\ dissertation of PW under the advisement of DG. 
DM obtained and reduced the VEGA observations.  All the authors contributed to the 
analysis, interpretation, and review of the results. 
\end{contribution}

\FloatBarrier
\facilities{CHARA(VEGA)}

\bibliography{diskms}{}
\bibliographystyle{aasjournal}



\end{document}